\documentclass[journal,draftclsnofoot,onecolumn,12pt]{IEEEtran}
\usepackage{amsthm,amssymb,graphicx,multirow,amsmath,color,amsfonts}%,ulem}
\usepackage[update,prepend]{epstopdf}
\usepackage[noadjust]{cite}
\usepackage{tikz}
\usepackage{subfig}
\usepackage{makecell}
\usepackage{gensymb}
\usepackage{longtable}
\usepackage{bbm} % for \mathbbm{1}
\usepackage{pdfpages}
\usepackage{balance}
\usepackage{multirow}
\usepackage{comment}
\begin{document}
\pagenumbering{gobble}
\graphicspath{{./Figures/}}
\title{
A Survey on Integrated Access and Backhaul Networks	
}
\author{
Yongqiang Zhang, Mustafa A. Kishk and  Mohamed-Slim Alouini
\thanks{The authors are with Computer, Electrical and Mathematical Science
and Engineering Division, King Abdullah University of Science and Technology (KAUST), Thuwal 23955-6900, Kingdom of Saudi Arabia. Email: \{yongqiang.zhang.2,\ mustafa.kishk,\ slim.alouini\}@kaust.edu.sa.} % remove the date for conference drafts
}

\maketitle

\begin{abstract}
Benefiting from the usage of the high-frequency band, utilizing part of the large available bandwidth for wireless backhauling is feasible without considerable performance sacrifice.
In this context, integrated access and backhaul (IAB) has been proposed by the Third Generation Partnership Project (3GPP) to reduce the
expenses related to the deployment of fiber optics for 5G and beyond
networks.
 In this paper, first, a brief introduction of IAB based on the 3GPP
release is presented.
Then, the existing research on IAB networks based on 3GPP specifications and possible non-3GPP research extensions are surveyed.
The research on non-3GPP extensions includes the integration of IAB networks with other advanced techniques beyond the currently defined protocol stacks, such as the integration of IAB to cache-enabled, optical communication transport, and non-terrestrial networks. 
  Finally, the challenges and opportunities related to the development and commercialization of the IAB networks are discussed.
\end{abstract}

% Note that keywords are not normally used for peerreview papers.

\section{Introduction}
Due to the dramatic growth in both the end-users population (e.g., smartphones and tablets) and demand for information service (e.g., video streaming and cloud computing), global mobile data traffic has skyrocketed in recent years~\cite{Soldani2015}.
By 2030, the global mobile traffic volume is predicted to increase 670 times compared with it in 2010~\cite{union2015imt}.
In order to deal with such exponential traffic growth, 
the density of base stations (BSs) is expected to substantially increase in the future.
The feasibility of denser BSs deployments (a.k.a., Network densification) was discussed in \cite{Bhushan2014} with the anticipated requirement of coping with the increasing traffic growth. 
As a promising approach for extending the cell area and meet the high capacity demand, network densification is aiming to provide a reliable access channel by reducing the distance from mobile users to BSs and increasing the spectrum reuse~\cite{agiwal2016next}.

However, in conventional network densification, one of the major drawbacks is the capital/operational cost of the optical fiber deployment for BSs. 
For instance, one-meter optical fiber deployment is estimated to cost approximately $100-200$ USD in a downtown area, nearly 85\% of the total expense results from the trenching and installing operations~\cite{willebrand2001fiber}.
In this context, there is an attractive implementation solution by  
using wireless backhaul instead of conventional wired-optical-fiber backhaul. 
Compared with wired-fiber backhaul, wireless backhaul can not only provide almost the same transmitting rate as optical fiber but also bring with considerable cost decline and more flexible/timely deployment (e.g., no intrusion)~\cite{Teyeb2019,Czegledi2020}.  

On the other hand, one of the objectives of 6G communications is to achieve worldwide connectivity~\cite{dang2020should}.
To this end, there is an emerging need for providing reliable communications service in remote and rural areas.
For the same reasons, it might be economically appealing
to use wireless backhaul rather than wired-optical-fiber backhaul to provide connectivity in such underserved areas.

In Long-Term Evolution (LTE) Rel-10, 3GPP first introduced the study item involved wireless backhaul, known as LTE relaying~\cite{3gpp10}. 
However, since spectrum resource in LTE is too valuable to be used for backhauling, it gained little commercial interest.
As several nationwide 5G mobile networks have been already launched, many other countries authorities have announced plans to grant 5G licenses for commercial use.
One of the key features of 5G is the usage of high frequencies transmitting carries (e.g., millimeter wave), with the aim to facilitate the prospective larger spectrum. 
The large available spectrum of the these bands empowers a dramatic improvement of transmission data rates, which support the carrier frequencies up to 52.6 GHz in millimeter wave (mmWave)~\cite{Rangan2014}.
Consequently, limited by physical properties, high-frequency carriers lead to limited coverage area and need higher density of BSs deployment.

The 5G new radio (NR) has been developing by 3GPP over the past few years, to meet the demands for higher system capacity, better coverage, and higher data rates~\cite{Parkvall2017}. 
Key design principles in 5G NR consist of ultra-lean transmission, advanced antenna technologies, and spectrum flexibility including operation in high-frequency bands~\cite{Lin2019}.
Benefiting from the wide bandwidth in 5G NR, operator is capable of partitioning the total radio resource into two parts for wireless backhauling and access, respectively. 
This technique is known as integrated access and backhaul (IAB) networks, which has recently attracted more research attention~\cite{dehos2014millimeter, Teyeb2019,Madapatha2020a}.
For 5G NR, I                    AB has already been standardized and recognized as a cost-effective alternative to the wired backhauling~\cite{3gpp}.
Compared to the LTE relaying, IAB NR has more potential to receive considerable industrial attention~
\cite{nokiawp,Ericssoniab,makki2019adaptive,wang2019methods}.
For example, in case of mmWave network, the shortage of narrow coverage will result in an urgent need to increase the density of BSs deployment. 
Consequently, there is a much higher demand for wireless backhauling. 
Meanwhile, by exploiting the much larger bandwidth, it is less expensive for operator to perform self-backhauling. 
Besides, a large number of antennas can easily be employed for mmWave-enabled BSs due to their small wavelength, which can enhance the signal directional gain and link reliability for backhauling.
\begin{figure}[!t]
\centering
\includegraphics[scale=0.5]{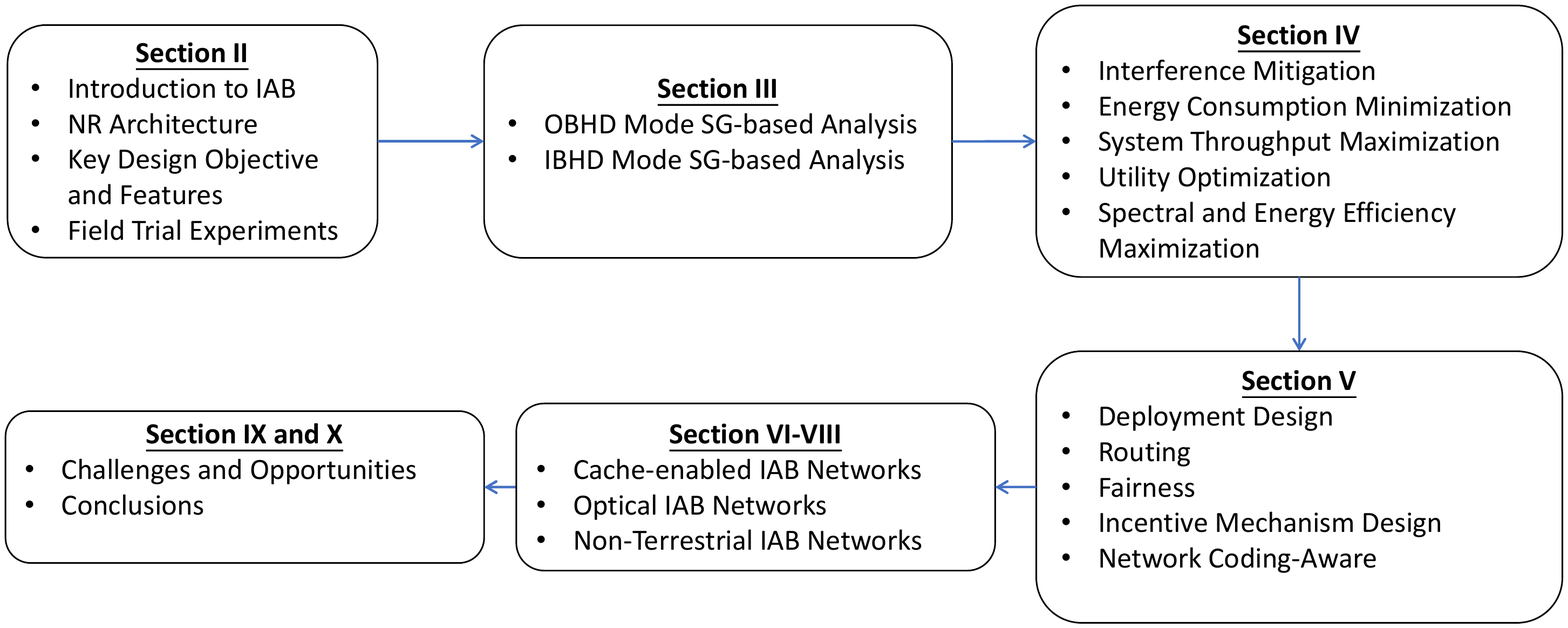}
\caption{Organization of the survey.}
\label{fig:org}
\end{figure}

Due to the increasing interest from both academia and industry, it is required to have a comprehensive overview of IAB networks for beginners in the related research field. In comparison to the vast amount of published work in the IAB networks, to the best of our knowledge, there is no survey that attempts to summarize the existing literature, which motivates this work.
As shown in Fig. \ref{fig:org}, the rest of the paper is organized as follows. 
In Section~\ref{sec:2}, we give a brief introduction of IAB architecture. 
The stochastic geometry-based analysis for IAB network is discussed in Section~\ref{sec:3}.
Section~\ref{sec:4} and~\ref{sec:5} focus on the resource allocation and scheduling research in IAB network.
Studies on the integration of IAB network with cache-enable network, optical communication, and non-terrestrial communication are surveyed in Section~\ref{sec:6}. Section~\ref{sec:7}, and Section~\ref{sec:8}, respectively. 
The challenges and opportunities of IAB are discussed in Section~\ref{sec:9}. 
The paper is concluded in Section~\ref{sec:10}.

\section{Integrated Access and Backhaul Technique}\label{sec:2}
\begin{figure}[!t]
\centering
\includegraphics[width=0.5\columnwidth]{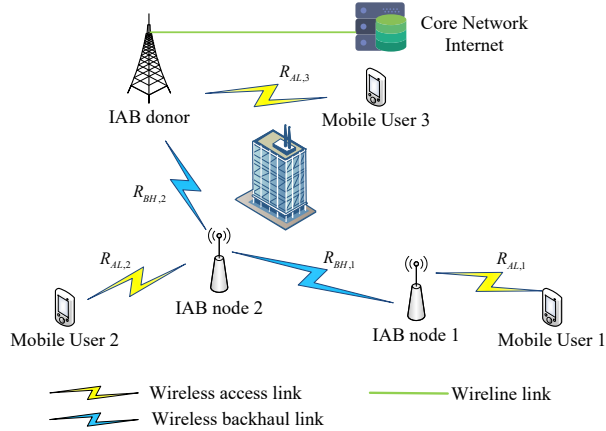}
\caption{IAB system architecture.}
\label{IABmodel}
\end{figure}
Over the course of the last 20 years, wireless backhauling in wireless networks has been studied extensively~\cite{Raza2011}.
However, there is no tight integration of access and backhaul in the studies of LTE backhaul networks, because only one single hop with a fixed parent BS is supportable for LTE relay, and strict resource partitioning is specified between access and backhaul~\cite{Polese2020}. 
In contrast, the IAB network offers a more flexible deployment solution without the heavy overhead of optical-fiber installation\cite{Dahlman2020}.  
Based on a wired connection to the core network, 
the IAB donor can provide communication access to mobile users
and wireless backhaul to IAB nodes.
IAB nodes are able to wirelessly provide network service access to the mobile user as well as the backhaul traffic. 
Therefore, IAB nodes can be regarded as a wireless relays for extending the coverage of an IAB donor.
This functionality is helpful for networks to achieve robust coverage performance when line-of-sight (LoS) propagation is blocked by environmental obstacles, such as buildings.
Fig.~\ref{IABmodel} provides a simple example of an IAB network, where $R_{AL}$ and $R_{BH}$ denote the data rates of the wireless access link and the wireless backhaul link. The achievable data rate from the mobile user to the IAB donor is determined by the minimum rate of the access link and the backhaul link. For example, in Fig.~\ref{IABmodel}, the transmission rate from the mobile user 1 to the IAB donor is  
 $\min\{R_{AL,1},R_{BH,1},R_{BH,2}\}$.

\vspace{0.5em}
\subsection{NR IAB Architecture}

With the objective to cope with the increasing demand for backhaul, 3GPP first proposed a study item on IAB in~\cite{3gpp15}. 
The physical-layer specification of IAB were completed at 2019, and higher-layer protocols and architecture was completed  in 3GPP Rel-16 at July 2020~\cite{3gpp}. 
Further enhancements (e.g., mobile IAB) has been carried out in 3GPP Rel-17, which is expected to be frozen in December 2020.
We will give a brief overview of the IAB NR in the following based on 3GPP specifications.

\begin{figure}
\centering
\includegraphics[width=0.6\columnwidth]{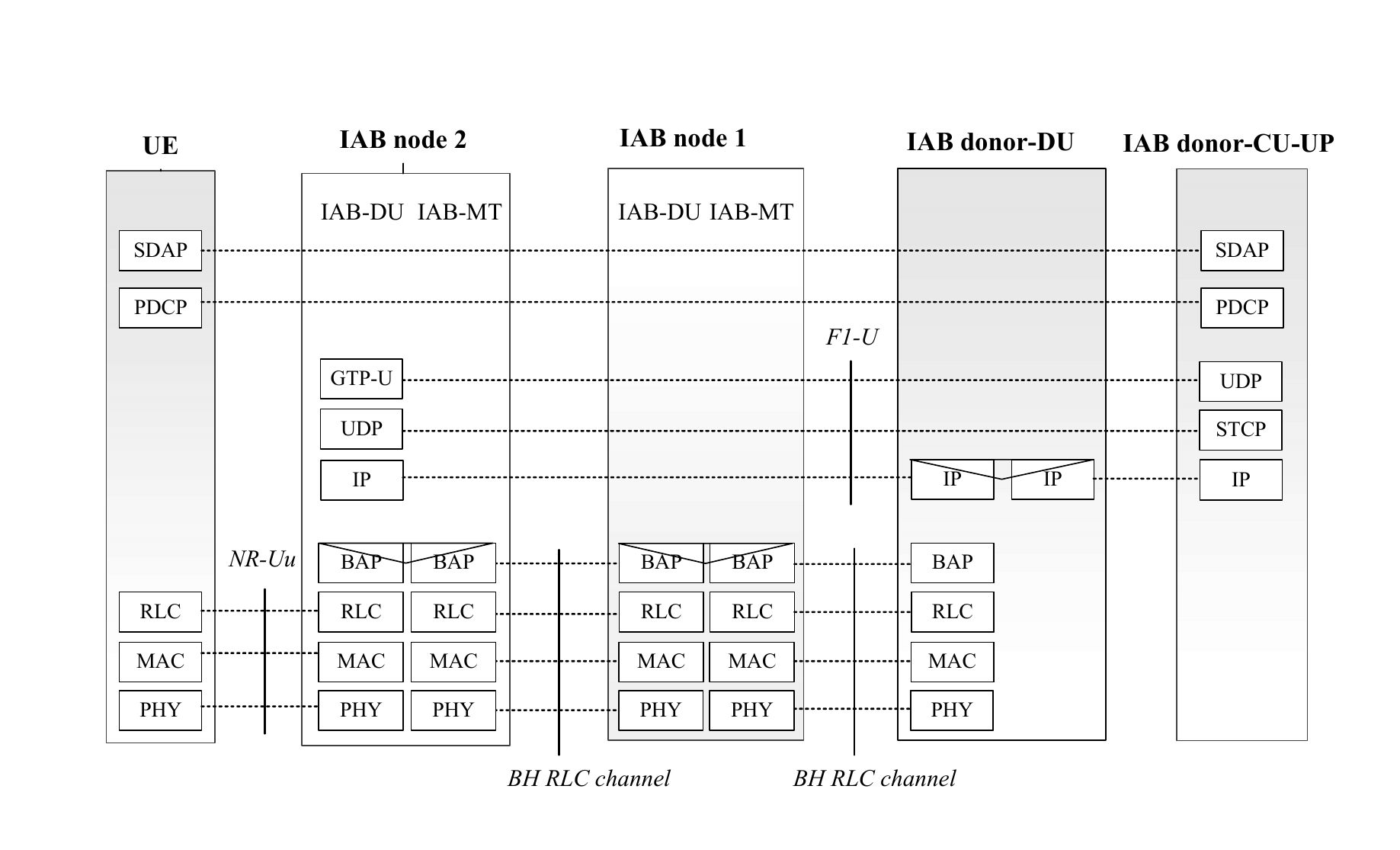}
\caption{IAB user plane protocol stack according to~\cite{3gpp}.}
\label{fig:U-Stack}
\end{figure}

\begin{figure}
\centering
\includegraphics[width=0.6\columnwidth]
{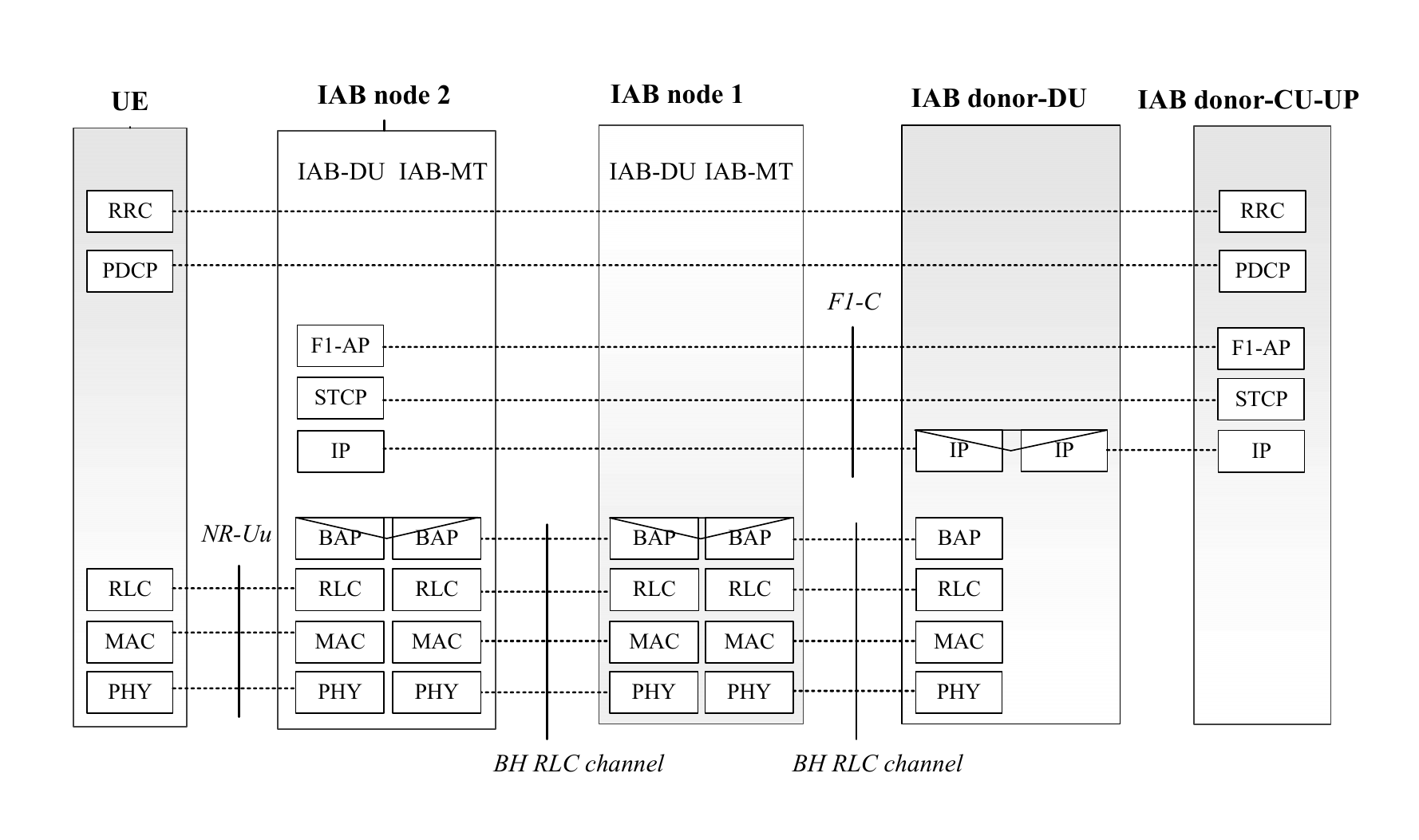}
\caption{IAB control plane protocol stack according to~\cite{3gpp}.}
\label{fig:C-Stack}
\end{figure}

Fig.~\ref{fig:U-Stack} and~\ref{fig:C-Stack} illustrate the IAB protocol stack for user plane and control plane, receptively~\cite{3gpp}. 
The overall architecture is based on the functionality split. 
In particular, the IAB donor consists of a central unit (CU) and no less than one distributed unit (DU).
The DU includes the Radio Link Control (RLC), Medium Access Control (MAC), and Physical layer (PHY) protocols.
Apart from Packet Data Convergence
Protocol (PDCP), the CU includes Service
Data Adaptation Protocol (SDAP) or Radio Resource Control (RRC) protocol in control plane or user plane, respectively. 
The interface between CU and DU is standardized as F1 interface, which defines the higher layer protocols.
The IAB donor connect the core network via non-IAB backhaul, and uses the IAB donor-DU wirelessly to serve UEs and the connected IAB nodes. 

IAB nodes comprise mobile termination (MT) and DU functionalities.
The IAB nodes rely on IAB for backhauling and provide service for UEs and IAB nodes via DU functionality.
The MTs act as normal devices and associate with the DU of parent IAB node or IAB donor. 
The message transmission is based on the lower layer functionality provided by the link between IAB node MT and its parent node DU.
Besides, IAB node can be backhauled to the IAB donor
through more than one intermediate IAB node, implying that  multi-hop backhauling is  supported in the IAB network. 

The lower three protocol stacks up to the RLC are known collectively as the NR-Uu interface. 
The middle three layers between Backhaul Adaptation Protocol (BAP) and PDCP provide the F1 interface for user plane (F1-U) and control plane (F1-C).

The BAP is a novel protocol defined by IAB, and is responsible for routing information packets from IAB donor to the target IAB node and vice versa. 
A typical IAB node has its own BAP address. 
For downlink (DL) scenario, the BAP layer of IAB donor first adds the BAP header to the information packets.
The BAP header comprises the routing ID for the destination BAP address and the path ID includes the path to the destination node.
In addition, a flag indicator included in the BAP header is determined 
by the packet type~(i.e., control plane or user plane)

Once the typical IAB node receives an information packet, the BAP layer will first check the routing ID in the BAP header.
If the IAB node is the destination node, the information packet will be forwarded to higher layers.
For example, the packet will be elevated to GTP-U or F1-AP when it is intended for UE served 
by the IAB node or it is a control plane packet for the IAB node.
Otherwise, the IAB node will deliver the packet to its DU and transmit the packet to next node based on the routing table.
\vspace{1em}
\subsection{Key Design Objective and Features}
The advantage of IAB lies in its supporting flexible and high-density deployments of NR cells without relying on the costly wired backhaul network deployment.
In addition, IAB is equipped for  a diverse range of deployment scenarios, including outdoor or indoor dense small-cell deployments and coverage extension.
\vspace{1em}
\subsubsection{Spectrum} 
Benefiting from the large bandwidth and small wavelength, mmWave IAB enables massive beamforming and using a portion of inexpensive bandwidth to do backhaul. 
Since high-band spectrum is generally organized as an unpaired spectrum, for operational purposes, 3GPP proposed that IAB enables wireless in-band and out-of-band relaying.
   
\begin{itemize}
    \item In-band relaying: Both access and backhaul links are simultaneously transmitted at the same frequency band. 
    \item  Out-of-band relaying: The access
    link and backhaul link transmissions are conducted in orthogonal channels. 
    That is, the access link transmission is conducted at a certain frequency band, and the backhaul link utilizes the remaining frequency band.
\end{itemize}
\subsubsection{Network Topology}
IAB enables multi-hop backhauling, affording a flexible range extension. flexible range extension~\cite{Teyeb2019}. 
In Rel-16, directed acyclic graph (DAG) based and spanning tree (ST) based multi-hop topology is supported. 
The end-to-end performance of IAB networks is strongly affected by the number of hops, the maximal number of relays that the IAB donor can support, and route selection~\cite{polese2018distributed}. 
As shown in~\cite{Gupta2019,Yuan2018, polese2018distributed}, with proper design of the route selection, IAB networks can utilize the advantage of multi-hop backhaul to improve the bottleneck performance.
%For example, the ergodic capacity for IAB networks was shown to be decreasing  as the number of hops increases in~\cite{Zhang2019a}.
%Although the problem of route selection in multi-hop wireless communication systems has been extensively studied over the last few decades, the optimization of routing and scheduling for multi-hop self-backhauled networks is still an open issue~\cite{Gupta2019}.
Furthermore, network topology adaptation and redundant connectivity are supported in IAB, enabling better backhaul performance and fast adaptation to handle link overloads or failures. 
\vspace{0.5em}
\subsubsection{Radio Link}
With the sharing of resources between access and backhaul, it may result in limitations on the end-user quality of service (QoS) (e.g. rate, latency). 
And the new interference results from the aceess/backhaul-to-backhaul/access needed to mitigate. 
One of the targets in radio link design is to ensure the QoS requirements of users are fulfilled even in a multi-hop setting.
Another important issue is that the deployment of IAB network should be transparent to UEs (i.e. no additional requirement for UE features/standardization).
\vspace{0.5em}
\subsection{Field Trial Experiments}
The advantage of mmWave IAB network was verified by a field experiment in \cite{Tian2019}. For instance, in \cite{Tian2019}, coverage ratio and user throughput in an out-door IAB network with time-division multiplexing were investigated. 
Their simulation results showed that compared to when an IAB node is not deployed, the coverage ratio can be improved to  approximately 78\% - 89\% with the deployment of only one IAB, and that a maximum coverage ratio gain of 46\% can be achieved.
The system throughput improved significantly by IAB node deployment, compared with non-IAB node deployment, up to $81.5\%$ throughput gain is obtained.
Since the wireless backhaul enables network densification without incurring additional fiber deployment, compared with wired optical fiber based backhaul, IAB can reduce a large portion of the total deployment cost tied to trenching and installation~\cite{Teyeb2019}. 
Moreover, IAB allows network operators to deploy small-cell BSs (SBSs) flexibly, which does not need digging or disruption of infrastructure ~\cite{Ericsson}.

\vspace{2.0em}

\section{Stochastic Geometry-based Analysis of IAB Networks}\label{sec:3}

Stochastic geometry (SG) has been regarded as a powerful mathematical tool for modeling, analyzing, and designing wireless
network~\cite{ElSawy2013}. 
In this section, we will introduce some papers with the integration of SG and IAB network.

\subsection{Out-of-band HD (OBHD) Mode}
In \cite{Singh2015}, Singh \emph{et} \emph{al.} developed a tractable mmWave IAB network analysis model, with the objective to characterize the rate distribution.
Moreover, the authors considered a more general setup with the mmWave IAB network co-existing with conventional ultra-high frequency (UHF) cellular network.
The proposed analytical framework was verified by two realistic environment topologies. 
It was shown that the density of BSs can improve the rate coverage probability drastically. 
Compared with the increasing of BSs density, the improvement of bandwidth
has a less positive effect on the rate of cell edge mobile users.

Because the bandwidth is shared between the access and backhaul links in the IAB-donor mode, the allocation of bandwidth plays an important role in network performance.
In~\cite{Saha2018}, the authors studied a setup that assumes a single macro-cell consisting of one anchored BS (i.e., IAB donor) surrounded by small-cell BSs (i.e., IAB nodes) while  the IAB nodes employed OBHD mode. 
By using SG tools to capture the locations of SBSs and mobile users as well as the load of BSs, Saha \emph{et} \emph{al}. presented an analytical framework to characterize the rate coverage probability accurately. For instance, the authors first assumed the bandwith of IAB donor is splitted into two parts: access bandwidth and backhaul link bandwidth. Further, Saha \emph{et} \emph{al}. investigated the impact of three different backhaul link bandwidth allocation strategies: 1) equal partition: all IAB nodes share the bandwidth equally;
2) instantaneous load-based partition: the backhaul link bandwidth allocated to an IAB node is proportional to its instantaneous load; and 3) average load-based partition: the backhaul link bandwidth allocated to an IAB node is proportional to its average load. 
The simulation results showed that the coverage probability $\mathbf{w.r.t}$ the bandwidth splitting ratio at the IAB donor behaves like a concave function (i.e., there exists only one optimal splitting ratio in all considered strategies). 
Moreover, in terms of both coverage probability and the median rate, the performance of three strategies can be sorted in a descending order as: instantaneous load-based $>$ average load-based $>$ equal.

In the follow-up work, Saha \emph{et} \emph{al.} considered a more general model setting which consists of multi macro-cells~\cite{Saha2019}, the locations of the fiber-wired macro-cell BSs (MBSs) were modeled by a Poission point process. The authors proposed two bandwidth allocation strategies: 
1) Integrated Resource Allocation (IRA): the bandwidth allocated equally for users; 
2) Orthogonal Resource Allocation (ORA): the MBS reserved a fixed fraction of bandwidth to allocate to its directly served users, and allocate the rest of the bandwidth for the backhaul link to SBSs proportional to the load at each small-cell. 
In terms of rate coverage probability, the simulation results showed that the IRA outperforms ORA, and the improvement increases with increasing the density of SBSs. 
The reason behind these performance trends is that the fixed fraction reserved bandwidth backhaul link cannot deal with the increasing backhaul load results from increasing density of IAB nodes.   

A semi-closed-form expression for the ergodic throughput of the two-tier IAB network was presented in~\cite{Zhang2018}.
Based on the proposed analytical model, the authors formulated and solved an optimization problem in order to maximize the number of offloaded mobile users, while satisfying the requirement for signal-to-interference-plus-noise ratio~(SINR). 
Numerical results showed that the proposed method can achieve two times the improvement for the average user and SBS throughput.
The joint uplink (UL) and DL coverage probability
under different resource allocation schemes in an mmWave-enabled IAB network was studied in \cite{Kulkarni2017}.
In particular, the authors investigated the impact of static and dynamic time-division duplexing (TDD) scheduling policies at the access link, as well as the synchronized or unsynchronized time alignment at the backhaul link.
Simulation results revealed that the combination of dynamic TDD and unsynchronized time alignment can achieve the best performance in terms of average DL and UL transmitting rate.
\vspace{0.5em}
\subsection{In-band FD (IBFD) Mode}
IBFD IAB node is the network framework in which IAB nodes can use the  same time and frequency resource block to conduct reception and transmission. 
Therefore,  there is no need to consider the bandwidth allocation in the IAB node operation. 
However, this mode will introduce self-interferenc~(SI), which is an undesirable transmission interference signal at the IAB- node's MT from the transmission from another IAB- node's DU to 
the user.
Since the distance of the interference source (DU) from the MT is significantly larger than that between the desired signal's source and the MT, SI is powerful enough to cause critical performance loss. 
Although numerous state-of-art SI cancellation technologies are available, residual self-interference (RSI) always remains~\cite{Duarte2010,Phungamngern2013,bharadia2013full,Kim2015}.

An analytical expression for coverage probability in a massive multiple-input multiple-output (MIMO) wireless self-backhaul networks was derived in~\cite{Tabassum2015} by using SG tools.
The considered two-tier network consists of a mixture of operating small-cells,
each small-cell is assumed to adopt in-band or out-of-band backhaul mode with a certain probability. 
Both the existence of self-interference and cross-tier  interference were considered for the signal-to-interference ratio (SIR) at mobile users. 
Simulation results revealed that implementing the considered mixture deployment consideration outperforms adopting only either in-band or out-of-band backhauling SBS option.

In~\cite{Sharma2017}, the authors developed a framework to analyze the coverage probability and average DL transmission rate for a two-tier IAB network, where the MBS (i.e., IAB donor) is OBFD-enabled, and SBSs are IBFD-enabled.
The trade-off between increasing interference and spectrum efficiency was investigated by the end-to-end analysis of the backhaul link and access link jointly.
Numerical results showed that the DL rate in the considered system setting could achieve nearly two times gain compared to a traditional TDD or frequency-division duplexing (FDD) self-backhauling network.
Meanwhile, due to the higher interference in IBFD mode, the coverage probability in the considered network is approaching half of its value in conventional TDD or FDD operation.
\subsection{Section Summary}
 The comparison of individual studies on SG-based IAB networks performance analysis is shown in Table~\ref{tab:TaxonomySec3}.
 The main objectives of these studies are developing analysis frameworks to characterize the coverage probability and the ergodic capacity in large-scale IAB networks accurately. 
Due to the existence of orthogonal bandwidth resource partitioning, the system performance of the OBHD IAB network is tied to the bandwidth partitioning ratio. This scheme can avoid the inter-interference between the access link and the backhaul link, with the sacrifice of efficient spectrum reuse. In order to enable efficient spectrum reuse, IBFD networks allow access and backhaul links to be transmitted in the same frequency band. However, the gains of IBFD transmission are degraded by the RSI, which is generated by the transmitter to its own receiver.

\section{Resource Allocation in IAB Networks}\label{sec:4}

\subsection{Interference Mitigation}
With the increasing mobile traffic and the existence of wireless backhaul hop, the number of transmission links in the IAB network is much larger than the past wireless networks that would introduce severe inter or intra interference at access and backhaul link.
Thus, the performance of interference management can be an important issue in IAB network design.
%%%%% Li2015 first IBFD backhaul mMIMO
In~\cite{Li2015}, the authors studied the average sum rate performance for three different duplex schemes in a massive MIMO-enabled IAB network.
For instance, a TDD based duplex, an IBFD mode, and an IBFD mode with interference rejection (IBFD-IR) were considered.
Based on zero-forcing (ZF) schemes, the authors formulated the beamforming matrix for UL and DL transmission in all the considered schemes. 
The simulation results showed that IBFD and IBFD-IR schemes outperform TDD schemes when the distance between IAB donor and IAB nodes exceeds some certain levels.
The performance for IBFD-IR is better than IBFD but the performance gap decreases as the distance between IAB donor and IAB nodes increases.
%%%%%%

%%%%% Ullah2016
Different from the case in~\cite{Li2015}, the authors in~\cite{Ullah2016} investigated the incorporation of beamforming into interference mitigation based on the usage of limited-feedback information.
In particular, for the DL transmission in an out-band IAB network, two different schemes based on antenna selection (AS) or quantized phase information (QCP) were proposed.
The corresponding beamforming and interference mitigation weight vector for AS and QCP were determined by the received signal power and SNR, respectively.
System-level simulation results showed that the combination of beamforming and interference mitigation can achieve nearly 9\% and 6.4\% compared with corresponding purely beamforming schemes.
The performance of QCP-based beamforming and interference mitigation was found to be the best among the considered techniques.	

%%%% 
In~\cite{Nakamura2019}, a comprehensive mechanism aiming at managing the out-band backhaul intra-channel interference for a mesh-architecture mmWave IAB network was proposed. 
The proposed mechanism consists of node placement, sub-channel alignment, and transmitting power control.
By solving a SIR maximization problem, the authors derived the optimal power allocation policy.
It was shown that the proposed method can reduce interference up to a distance of 250 m away from the wired-backhaul node.

Flexible TDD based resource scheduling is able to manage the frequency-time resources more efficiently. 
In \cite{jayasinghe2020traffic}, an iterative beamformer design for a TDD based IAB network with one IAB donor and multiple IAB nodes was proposed. 
For a given time slot, IAB donor and IAB nodes operate in different UL/DL modes. 
By considering the traffic dynamics at each node and assuming each user and access point (AP) has a specific UL/DL queue, Jayasinghe~\emph{et} \emph{al.} formulated and solved an optimization problem with the objective to minimize the weighted $l_p$-norm queue
minimization of the UL and DL users during two time slots.
By comparing the proposed method with the half-duplex (HD) IAB system, the results showed that the performance of the proposed scheme is better for all considered traffic arrival rates.

\subsection{Energy Consumption Minimization}
In~\cite{Prasad2017}, the authors proposed a coordinated mechanism to minimize network power consumption in an IAB network.
By utilizing the information of QoS in the whole network,
the controller cascaded with MBS can determine and inform which BSs or mobile users should enter to sleep mode along with the possible sleep time by discontinuous reception configuration (DRX) . 
By simulation based on realistic network parameters, the presented mechanism is able to achieve up to 50\% power saving in comparison with LTE. 
The problem to find optimal time allocation and power control for the mmWave IAB network was addressed in \cite{Meng2018}. The authors considered a multi-hop transmission system model, with one MBS and mesh connected SBSs.
Each hop equipped with a status indicator to guarantee the hop transmitting priority. 
By formulating an optimization problem with the objective to minimize the overall energy consumption under throughput constraints, Meng  \emph{et} \emph{al.} first established space/time-division multiple access groups, and then proposed a heuristic
algorithm to derive the time allocation and power control policies. 
The proposed scheme was shown to be achieving a considerable performance boost compared with the STDMA scheme~\cite{zhu2016qos} and the maximum QoS-aware independent set~(MQIS) scheme~\cite{qiao2012stdma}.

The energy consumption minimization problems in full-duplex (FD) IAB network was studied in~\cite{Korpi2018} and~\cite{Lei2018}.
Korpi \emph{et} \emph{al.} proposed a resource allocation scheme to minimize the energy consumption of a two-tier massive MIMO-enabled IAB network under QoS constraint in~\cite{Korpi2018}. %%%%%%%%%%%% 
In addition to the FD IAB scheme, a closed-form expression for the optimal transmitting power for HD and hybrid FD schemes were derived. 
In order to compare the performance of a three considered schemes, the authors further derived feasibility boundaries for the DL and UL rate requirements in closed-form.
It was shown that both FD and hybrid relay schemes are not able to guarantee the QoS constraints even as the transmitting power approaches infinity.
Numerical results reveal that the FD schemes can achieve the lowest energy consumption among the three considered schemes.
With the consideration of using non-orthogonal multiple access (NOMA) to enhance the spectrum efficiency,  Lei \emph{et} \emph{al.} presented an effective algorithm
to minimize the energy consumption for a two-tier NOMA-enabled FD IAB network based on fixed-point iteration in~\cite{Lei2018}.
Numerical results showed that the performance gap between the proposed method and the orthogonal multiple access (OMA) based networks increases as the transmission rate requirements increases.
\vspace{3em}
\subsection{System Throughput Maximization}
%Since severe co-tier interference and cross-tier interference resulted from neighboring base stations in ultra dense IAB network, employing full-spectrum
%reuse or other static schemes will be less efficient.
The rate of the mobile user associated with IAB node is determined by the minimum rate between the backhaul link at this IAB node and the access link, which is obviously sensitive to the bandwidth partitioning strategies. 
The spectrum allocation can significantly influence the user's data rate, as analyzed in \cite{Saha2019}.
Lei~\emph{et} \emph{al.} derived the optimal bandwidth allocation policy for a two-tier IAB network, by solving a mixed-integer nonlinear programming problem with the objective to maximize the sum log-rate, with the usage of deep reinforcement learning (DRL) to derive the optimal solution  in~\cite{lei2020}.
Benefiting from the DRL, it is tractable to obtain the optimal allocation policy for the large-scale time-varying IAB network.
Compared with the full-spectrum reuse strategy, simulation results showed that the proposed method can achieve considerable performance gain especially when the density of IAB networks is large.
With the consideration of UL-DL  and access-backhaul link transmission rate requirements, the joint UL and DL resource allocation mechanism for a TDD IAB-network was studied in~\cite{Liu2018a}. 
In this mechanism, the optimization problem aims at maximizing the overall throughput and was decomposed into two subproblems on time slot orchestration and sub-band scheduling. 
Numerical results showed that total system throughput improves significantly compared with conventional Round Robin (RR) and Proportional Fair (PF) mechanisms. 

As for in-band IAB network, the trade-off between self-interference and spectrum efficiency was investigated in~\cite{Lagunas2017,Zhang2020b,Korpi2016,Siddique2017}.
The authors in~\cite{Lagunas2017} obtained the optimal power allocation policy by solving the DL sum rate maximization problem in an IBFD IAB network.
Based on the standard optimization method, a closed-form expression for optimal transmitting power was derived in~\cite{Korpi2016} by solving the optimization problem with the objective to maximize the network sum rate of DL and UL.
The authors considered both the IAB donor and IAB node are working in IBFD mode, where the mobile users adopt HD mode. 
Simulation results showed that the proposed solution can achieve the best performance when UL and DL rate requirements are close.
In \cite{Zhang2020b}, the optimal spectrum and
power allocation policies were derived by  sequential convex programming (SCP) method. 
The authors considered a mmWave in-band IAB network with one IAB donor surrounded by multiple IAB nodes, where partial users could reuse the backhaul bandwidth resources.
The formulated problem was under the heterogeneous user's DL transmission rate requirements. 
Compared with the simple power allocation optimization, the proposed algorithm can achieve a nearly 25 \% gain in terms of system throughput.
Similar to~\cite{Korpi2018}, the authors in~\cite{Siddique2017} investigated the sum-rate maximization problem in IBFD, OBFD, and hybrid IBFD/OBFD backhauling network. 
By using the convex optimization method and bisection search algorithm, the authors derived the solutions for optimal spectrum allocation policies for three considered scenarios.   
With two heuristic methods based on received signal power,
three distributed algorithms and coverage probability were derived for corresponding backhauling schemes.
It was shown that IBFD outperforms OBFD when self-interference cancellation exceeds a certain level.
Moreover, IBFD tends to allocate more spectrum resources for the SBSs close to MBS while OBFD is just the opposite.

Using tools from game theory, the authors in~\cite{Liu2018} obtained the joint optimal power control and transmission slot allocation policies for a two-tier mmWave IAB network.
For the formulated non-cooperative game with the objective of sum rate maximization, the authors first proved the existence and feasibility of the Nash equilibrium then designed a centralized resource allocation algorithm as well as a decentralized algorithm based on the functionality splitting architecture at IAB nodes to solve it.
Simulation results showed that the proposed algorithms can achieve at least 17.94\% improvement compared with pure optimization for power control.
The joint optimal backhaul and access link resource allocation policies for a two-tier IAB network were studied in~\cite{Lashgari2017}.
In order to maximize the overall DL data rate,
the problem was formulated into a Stackelberg game form. The wired-backhaul MBS plays the role of the leader, while the SBSs act as the followers. For instance, the optimal sub-carriers allocating policies for MBS and SBSs were derived from the leader's and followers' optimization problems, respectively. 
An optimal power control strategy was obtained from the follower's problem. 
Numerical results proved that the presented algorithm can improve the performance up to  14.2\% and 24.9 \%.
%%%
By integrating game theory tool and reinforcement learning (RL),
the authors in~\cite{Blasco2013} presented a game-theoretic leaning mechanism to maximize the UL sum rate in an in-band IAB network.  
The proposed mechanism is supportable for self-organizing, which means that the MBS enables throughput balancing in a decentralized way.
In comparison with non-IAB mechanism, the proposed mechanism can achieve up to 40\% performance improvement.

With the objective to maximize the weighted sum-rate in a mmWave IAB heterogeneous network (HetNet), the joint optimal strategies for transmitting power allocation, time splitting ratio, mobile users association, and beamforming design were proposed in~\cite{Kwon2019}. 
Time splitting manner is considered between backhaul and access links. 
The access links design includes the mobile users association and beamforming design by using limited channel state information (CSI). 
By simulation, the performance of the proposed limited feedback hybrid beamforming scheme was shown to be able to approach the performance of the digital beamforming scheme with full CSI. 
The spatial multiplexing gain increases up to the number of radio frequency (RF) chains for IAB node.

The DL sum rate maximization problem in an enhanced hybrid IAB network was studied in~\cite{Ni2019}. 
For instance, with the aim to extend the coverage of IAB nodes, the authors first considered that the access link is working in sub-6 Hz while the backhaul link is working in mmWave, then proposed a hybrid precoder to improve the backhaul transmission rate.
Numerical results revealed that the performance of the presented precoder is close to conventional block diagonalization precoder, with a significant decline for the required number of the RF chains.
In~~\cite{Pu2019}, the authors considered a network with multi-antenna IAB nodes, and the transmission link can be allocated with an orthogonal sub-channel when it suffers from severe interference.
By introducing the penalty function and factors, Pu~\emph{et~al.} first transformed the sum rate maximization problem into 0-1 integer programming without any inequality. 
Further, a resource allocation algorithm based on Markov approximation with polynomial time complexity was proposed.
The performance evaluation showed the sum rate for the proposed scheme increases faster than OBFD scheme as the number of antennas increases.

In~\cite{Yang2020}, the authors addressed the sum rate maximization problem in a NOMA-enabled out-band IAB network, where a typical mobile user is only assumed to be served by two SBSs.
Two decoding strategies based on different SIC decoding order
were considered at DL access and backhaul links. 
By using convex-concave procedure (CCP) method, the authors derived the optimal power allocation policies for UL and DL transmissions.
It was shown that decoding better channel quality signals first outperforms the other decoding strategy.
Compared with OMA, the performance improvement of considered NOMA-enabled network increases as the number of mobile users increases.
With the consideration that mobile users can be served by a cluster of SBSs cooperatively, the authors in~\cite{Chen2018} addressed the joint beamforming and SBSs clustering problem for IBFD IAB network.
In order to maximize the overall DL transmission rate,
a manifold optimization problem was formulated under the transmitting power constraints. 
By using Riemannian optimization technique, a heuristic algorithm was presented to derive the optimal policies for the SBSs clustering and beamforming vectors.
Extensive simulations showed that optimal cluster size is related to the transmitting power constraint for the IAB donor.
The extension of~\cite{Chen2018} from full CSI to partial CSI is presented in~\cite{Chen2019}.
To this end, a stochastic successive lower-bound maximization algorithm and a deterministic algorithm with lower time complexity was proposed to solve the modified sum-rate maximization problem. 
Moreover, the performance gap between~\cite{Chen2019}~and~\cite{Chen2018} was shown to be decreasing as the cluster number increases.  
\vspace{3em}

\subsection{Utility Optimization}
The maximization of network utility with interference management in the DL transmission of a mmWave IAB HetNet was considered in \cite{vu2017joint}. 
The authors developed a Lyapunov optimization framework to decouple the primal optimization problem and used convex optimization and successive convex approximation method to derive the optimal solution for user association,  beamforming design, and power allocation, respectively. 
Simulation results revealed that the proposed scheme can achieve 5.6 times gain in cell-edge users throughput compared with un-optimized user allocation scheme with a density of SBSs equal to $350/{\mathrm{Km}^2}$. 
A more general setting for multi-hop multipath scheduling is contemplated in \cite{Vu2019}. Vu \emph{et} \emph{al.} formulated a network utility function under low latency and network stability in order to achieve Ultra-low latency and reliable communication (ULLRC). 
A RL-based algorithm and a Lyapunov optimization algorithm were proposed to derive the optimal solution for path selection and rate allocation, respectively. 
Compared to \cite{vu2017joint}, it was shown in the simulation that the proposed scheme can provide reliable communication with a guaranteed probability near to 100\% and latency reduction at least 50\%.

Wireless network virtualization empowers the resources at network infrastructure provider (InP) to be sliced into several virtual parts, which can be shared by multiple service providers (SPs)~\cite{liang2014wireless}.
%%%% Chen2016a
In~\cite{Chen2016a}, the authors investigated the application of wireless network virtualization technique into a FD IAB network.
They assumed both MBS and SBS from multiple InPs can be virtualized and shared by mobile users from different mobile virtual network operators (MVNOs).
To solve the utility maximization problem with the aim to maximize  all the profit of MVNOs and  cost of resource consumption, a two-stage iterative algorithm based on convex programming was proposed.
Numerical results showed that the proposed algorithm can achieve convergence within 15 iterations and significant average throughput improvement for small-cells. 
The authors in~\cite{Tang2018} addressed the time-average utility maximization for an IAB network with wireless network virtualization.
In particular, they considered the requirements for network queue stability, average throughput of different SPs, and the backhaul link capacity.
A Lyapunov optimization-based algorithm was proposed to find the optimal strategy for bandwidth portioning ratio, user association, and percentage of resources allocated by the associated BS. 
In comparison with the traditional networks without virtualization techniques, the proposed scheme can result in at least 60\% increment of average total utility.
\vspace{0.5em}
\subsection{Spectral and Energy Efficiency Maximization}
In~\cite{Chen2016c}, the authors considered integrating energy harvesting (EH) technique into an IBFD IAB network with the aim to improve energy efficiency (EE).
The MBS is equipped with massive MIMO antennas array, and the SBS is equipped with a multi-antenna array and can harvest energy from renewable resources.
The authors first proposed a precoder to mitigate the self-interference and cross-tier interference, then used CCP method to obtain the optimal power control and mobile users association policies by solving the energy efficiency maximization problem.
Compared with the conventional OBFD network, it was found that the proposed method can dramatically improve EE for dense SBSs deployment.

The spectral efficiency maximization problem in the IAB network was studied in \cite{Imran2014,Chen2016b}.
%%Imran2014
In particular, the joint optimization of backhaul and access links with the aim to maximize spectral efficiency (SE) in OBHD IAB networks was studied in~\cite{Imran2014}. 
By using a sequential quadratic programming (SQP) algorithm, the optimal antenna tilt angle deployment policies were derived.
The proposed schemes can achieve up to 50\% performance improvement compared with fixed-tilt deployment.
%%Chen2016b one user per SBS 
With the consideration of IBFD mode, Chen \emph{et} \emph{al.}
addressed the SE optimization problem which under DL backhaul capacity requirements in~\cite{Chen2016b}.
After the transformation of the primal optimization problem, the optimal power allocation policy was derived using SCP-based algorithm.
IBFD IAB network was shown to be able to bring about 25\% SE performance improvement.
\vspace{0.5em}
\subsection{Section Summary}
 A comparison of the studies addressing the resource allocation problems for the IAB networks is shown in Table~\ref{tab:TaxonomySec4}. 
The related literature aims at finding optimal policies to improve the system performance, including of interference mitigation, energy consumption minimization, throughput maximization, utility optimization, and the maximization of spectral and energy efficiency.
A common drawback of all mentioned methods in this section is that the deployment of the IAB nodes was disregarded. With the realization of network densification, the QoS of UEs is highly dependent on the lengthes of the transmission links, especially when the BSs are adopting high frequency signal carriers. Besides, the proper design of scheduling policies are important to reduce the UEs dissatisfaction in IAB networks.
This issue is addressed in the following section which is targeting scheduling in IAB networks.

\section{Network Deployment and Scheduling in IAB Networks}\label{sec:5}
Since the wireless links in the network are implemented by sharing the communication resources, the flexible deployment of the IAB network brings an emerging demand for efficient scheduling to realize high throughput and low latency while guaranteeing users fairness. The wireless communication resources scheduling consists of the user scheduling and the link scheduling~\cite{Pathak2011,Ge2018}. 
For instance, user scheduling is aiming at exploiting the multi-user diversity by the proper selection of users for transmissions in different time slots, and the goal of the link scheduling is to achieve conflict-free feasible transmission schedule based on the estimation for the interference conflicts between the different links with different transmission demands~\cite{Hajek1988}. 
Due to the complexity of network topology in IAB networks, in addition to classical scheduling techniques (e.g., Proportional fair (PF) and backpressure algorithms), researchers have proposed some novel centralized or distributed schedulers based on RL, game theory, and matching theory~\cite{Zhang2020c,Chaudhry2020,Yuan2018}.   
This section surveys current research related to the design of network deployemnt scheduling policies in IAB networks. 
\vspace{1em}

\subsection{Deployment Design}

%\textcolor{red}{Since the usage of higher carrier frequency will result in higher pathloss and penetration loss, the coverage area and capacity of IAB networks are limited by both the number and locations of BSs.
%Hence, the designing of BSs deployment is an important issue in IAB networks.}
Due to the higher pathloss and penetration loss, the coverage area of the network operated in high frequents is limited by both the number and locations of BSs.
In \cite{Bonfante2018}, the authors investigated the ad-hoc deployments in a massive MIMO-enabled IAB network, where the SBSs were positioned in proximity to mobile users. 
In~\cite{Wainio2016}, a self-optimizing deployment framework for an IAB mesh network was developed.
Based on the neighbor discovery, the proposed framework enables autonomous deployment for the newly added nodes.
Taking into account the height of nodes, the authors investigated the joint resource allocation and node deployment problem for a MIMO-orthogonal frequency-division multiple access (OFDMA) based IAB network in~\cite{Lai2020}.
Since only IAB donor is connected to the core network via wired fiber in IAB network, the location of IAB donor in a set of BSs can determine the quality in multi-hop wireless backhauling.
A genetic algorithm (GA) in combination with the K-means clustering method to select the location of IAB donor to maximize the backhaul capacity was proposed in \cite{Raithatha2020}. 
Through extensive Monte Carlo simulations, the authors evaluated the performance of this algorithm in comparison with the conventional genetic algorithms and K-means clustering in terms of the average number of hops and backhaul network capacity at different node densities.
Simulation results showed that the proposed algorithm can achieve at least 20.2\% and 19.8\% a performance improvement in the average number of hops and backhaul capacity, respectively.

The trade-off between deployment cost and network performance brings  more challenges in deployment design in IAB network. 
%%%
In~\cite{islam2017integrated}, the authors formulated a mixed-integer linear program problem with the objective to minimize the cost of required fiber-wired BSs deployment and used the branch-and-bound algorithm to obtain the optimal solutions of deployment, routing, and resource allocation on each flow in a single-tier IAB network.
Simulation results showed that IAB is able to reduce the amount of fiber-wired BSs deployment significantly.
Another interesting conclusion is that there were only 6.5\% links larger than the specific noise tolerance when the optimization problem formulation did not consist of interference from inter-links.
The BSs deployment problem for a two-tier IBFD IAB network was studied in~\cite{Rezaabad2018}. 
With the aim to minimize deploying cost and maximize coverage,
the optimal number of IAB donors and IAB nodes were derived from a non-dominated sorting genetic algorithm.
Numerical results confirmed the trade-off between BSs deployment costs and coverage.
For realistic network topology, authors showed that the optimal deployment solution can be derived from the Pareto front of the multi-objective optimization problem according to operator policy.
\vspace{0.5em}
\subsection{Routing}
With the introduction of wireless backhaul links and the support of multi-hop backhaul, the design of routing policies in the IAB network should be able to scale together with the increasing complexity.
The end-to-end ergodic capacity for an IAB network was shown to be decreasing as the number of hops increases in~\cite{Zhang2019a}.
%%%%%Hui2013
In order to maximize the overall sum rate as well as minimize the average latency, the joint route selection and radio resource allocation optimization for an indoor IAB network was studied in~\cite{Hui2013}.
Based on the virtual-network method, the design of routing and radio resource allocation among different links can be optimized independently.
As expected, by ray-tracing simulations, the number of IAB donors has a positive impact on the performance of the proposed centralized scheme.
%%%%%Favraud2017 
The authors in~\cite{Favraud2017} considered an outdoor in-band mesh-architecture IAB network. 
Under the constraints of QoS, the centralized and semi-centralized joint routing and power control mechanism were presented.
This mechanism is realized by a logical controller which
manages the network across the MAC layer.
It was shown that the proposed mechanism does not need intense modifications on current standards while providing robust performance for real-time flows.
  
%%%%islam2018investigation
The optimal flow link resource allocation and routing policies to maximize the geometric mean of users in a single-tier multi-hop network mmWave IAB network is proposed in \cite{islam2018investigation}. 
The authors investigated different routing patterns, and found that the only top 20 percentile user rates in IAB with RSRP-based ST outperforms IAB with mesh. 
The suboptimal allocation of the subcarriers and the transmitting power are derived by the dual-decomposition method, and a systematic approach to determine the optimal node locations was developed. 
Simulation results revealed that the performance gap between the optimal and the proposed suboptimal method is negligible.
In~\cite{Lukowa2018}, the authors addressed the dynamic scheduling problem for a mmWave DAG network, where BSs are equipped with multiple antenna panels.
By considering flexible access/backhaul TDD and dynamic route selection, a heuristic based centralized scheduling scheme was presented.
Both median and $5^{th}$ percentile of UL and DL rate for the proposed scheme was shown to be improved significantly by system-level simulation.
%%
%The authors in~\cite{Miao2015} developed a method of routing in an mmWave IAB network.
%A multi-hop backhaul path selection was proposed by utilizing the bidirectional beam alignment. 
%Dynamic routing was supported by the flexible wireless backhaul establishment. 

%%%% 
Different from the centralized solution in~\cite{Hui2013,islam2018investigation}, four distributed greedy path selection strategies: 1) Highest-quality-first (HQF), 2) Wired-first (WF) policy, 3) Position-aware (PA) policy, and 4) Maximum-local-rate (MLR) were studied in~\cite{polese2018distributed}.
The HQF and MLR policies are implemented by selecting the node with the best SNR and achievable rate, respectively.
As for the WF policy, the IAB node prefers to connect with the  reachable IAB donor if the SNR exceeds a given threshold.
Otherwise, the IAB node will connect with the node with the highest SNR.
The PA policy is based on the position information, the IAB node will connect with the highest SNR node selected from the candidate nodes which are closer to the target IAB donor than current IAB node. 
Moreover, in order to maintain a small number of hops for the proposed HQF, WF, and MLR policies, the authors introduced two wired bias functions which are polynomial and exponential in the number of hops, respectively.  
The wired bias function is applied to the SNR of the wired IAB donor, and takes the value which is larger than the fixed tolerable SNR level when the number of traveled hops reaches the threshold.
Simulations revealed that aggressive bias function is able to decrease the number of hops needed to connect to an IAB donor.

%FD self-backhauling allows exploiting the radio spectrum used by the radio access network (RAN) for backhaul links as well as access links concurrently. 

As shown in~\cite{Zhai2020}, mesh-architecture IAB network can achieve 6.70\%$\sim$40.56\% overall throughput improvement compared with DAG-architecture IAB network.
The authors in~\cite{GomezCuba2020a} addressed the joint link scheduling and power allocation problem in a mesh-architecture IAB network with multi-user MIMO (MU-MIMO) and elastic traffic arrival process.
To maximize the overall throughput, a modified backpressure algorithm and an adaptive congestion control algorithm were proposed to obtain the optimal policies for link connectivity decision, power allocation, and traffic allocation.
The proposed method was proved to achieve a stable performance by stochastic analysis.
However, the study in~\cite{GomezCuba2020} is based on the perfect link directivity, which ignores the inherent interference.
In this context, the authors considered using a simulated annealing heuristic algorithm to solve the overall throughput maximization problem, and investigated the impact of minimal feedback and sophisticated beamforming.
Simulation results showed that a more sophisticated beamformer enables the proposed algorithm to achieve sum rate performance closer to interference-free scenario.

%Both \cite{Kwon2019} and \cite{Ni2019} are only considered the DL transmission, however, it is known that one of the advantages of IAB compared to traditional relay network is that IAB enables UL and DL transmission at the same time. 

In order to obtain a tractable optimal solution for the routing in harsh propagation and dynamic environment at mmWave IAB networks, RL with the promising capacity to deal with a massive number of parameters was applied in~\cite{Gupta2019,Zhang2020c}.
By utilizing full or partial knowledge of the network state, the authors in~\cite{Gupta2019} proposed two corresponding routing policies based on the integration of deep deterministic policy gradient (DDPG) algorithm into RL.
The optimal routing policies are aiming at minimizing the transmission delay by allocating the proper time slot for each link.
Numerical results show that the proposed method can achieve at least 2.6 times gain compared with max-min scheduler~\cite{Kulkarni2017} and backpressure scheduler~\cite{Tassiulas1992}.
%%%% Zhang2020c
In \cite{Zhang2020c}, the authors introduced a RL-based approach to maximize throughput in a mmWave IAB network.
The routing policy is determined by the column
generation (CG) method and the flow control policy was derived by RL-based algorithm.
Further, two schedulers for off-line and on-line model framework were proposed, respectively.
Compared with the purely CG-based method and random pattern selection method, the proposed method can achieve at least 10\% performance improvement.

%%% Chaudhry2020
It is clear that the wireless backhaul capacity is related to the number of simultaneous transmitting links.
The authors in~\cite{Chaudhry2020} proposed a routing mechanism to optimize the average number of simultaneous transmissions and hops.
The problem of maximizing the average number of simultaneously transmitting links is similar to the minimum coloring problem, 
a heuristic algorithm was proposed to solve this problem.
The average number of transmitting hops is determined by  Dijkstra's shortest path algorithm.
Numerical results showed that the backhaul network capacity increases as the node density increases or the interference range decreases. 
In~\cite{Ortiz2019}, the authors proposed a semi-distributed learning-based algorithm to minimize the end-to-end latency while enhancing the robustness against network dynamics in a two-tier IAB network.
The considered network dynamics include load imbalance, channel variations, and link failures.
By formulating a Markov decision process to capture the routing and resource allocation decisions, a learning algorithm without the requirement of prior information was developed.
Simulation results showed that the proposed algorithm achieves at least 1.8 times throughput gain compared with the RL benchmark.
\vspace{0.5em}
\subsection{Fairness}
From the user's point of view, the service provider needs to focus on fairness to meet their satisfaction requirements.
In general, the fairness in wireless networks is attributed to the resource sharing. An unfair resource allocation among different users will result in resource starvation, wastage or redundant allocation. For different researchers, it is rather difficult to agree on a single definition of fairness since it is subjective~\cite{Shi2014}.
Based on the measurability, the fairness measures can be classified as quantitative or qualitative. The Jain's index~\cite{jain1984quantitative} and the Shannon entropy~\cite{shannon2001mathematical} are the two most common quantitative measures of fairness, and the qualitative measures of fairness can be characterized by  $\max$-$\min$~\cite{radunovic2007unified} and proportional fairness~\cite{kelly1997charging}. 
By optimizing a specific type of fairness measure, the fairness performance in IAB networks was shown that can be improved in~\cite{Yuan2018,Siddique_2015,zhang2020co,Goyal2017}.
In this subsection, we survey the scheduling studies in the IAB network intending to guarantee users fairness. 
%%%
%The performance comparison of the self-backhaul network of HetNets with self-backhauled small cells (SBSCs) relative to those with wired backhaul based small cells (WBSCs) was investigated in \cite{Andrews2017}. 

In~\cite{Yuan2018}, Yuan \emph{et} \emph{al.} addressed the maximum
throughput fair scheduling (MTFS) problem in a TDMA mmWave HetNet with one multiple-RF-chains-equipped IAB donor and multiple singe-RF-chain-equipped IAB nodes.  
The topology of the considered system is represented by a directed graph, corresponding to an adjacency matrix.
A two-step approach was presented to solve the MTFS problem:
first, maximize the max-min throughput and then find the optimal scheduling solution to maximize network throughput for the given max-min throughput.
Based on matching theory and ellipsoid algorithm, the authors first proposed an optimal MTFS algorithm with polynomial time complexity.
Next, they developed an edge-color approximation algorithm with improved runtime efficiency.
Simulation results showed that the proposed optimal algorithm could converge with a few minutes for over 200 IAB nodes. 
The edge-color approximation method can be 5 to 100 times faster with a performance decline up to 20\%. 

%%%%%% Siddique_2015
The authors in~\cite{Siddique_2015} investigated the scheduling problem in a hybrid IAB network.
The considered network enables an IAB node to switch its operating mode between HD and IBFD.
Analytical expressions for DL and UL transmission rates for users were first derived.
Based on the overall fairness of users, IAB node can select the mode which contributes the largest value.
For dense IAB node deployment scenario, simulation results showed that IAB nodes prefer to select IBFD mode to improve users' fairness.
PF scheduling was proved to be not suitable for multi-hop network in~\cite{Andrews2017}.
The impact of applying a modified PF scheduling in a multi-hop IAB network was studied in \cite{zhang2020co}. 
By adding a weight related to the numbers of users served by IAB node, the authors proposed a weighted proportional fair (WPF) algorithm.
Moreover, an IAB-aware flow control algorithm aiming at improving the system throughput and mitigating congestion were given. 
Compared with the original PF integrated with traditional end-to-end flow control, the proposed scheme can result in a 30\%  and 2\% increment of fairness index and system throughput, respectively.
Under a strict fairness requirement, the joint flow control, user association, and power control for an in-band IAB network was studied in~\cite{Goyal2017}.
The optimal power control and flow control policies were derived by using a backpressure algorithm and geometric programming, respectively. 
In comparison with the out-band IAB network, the throughput at the in-band IAB node was shown to be doubled.
\vspace{0.5em}
\subsection{Incentive Mechanism Design}
As a promising approach to address the multi-objective optimization problems in wireless networks, the application of economic and pricing models in the wireless network has attracted much attention from the research community~\cite{Luong2019}.
The trade-off between the revenue of IAB nodes and mobile users
satisfaction in an IBFD network was investigated in~ 
\cite{Rahmati2015}.
The authors proposed a price-based scheme by using Stackelberg game theory.
For instance, IAB nodes act as the leaders, while their mobile users act as the followers. 
The mobile user can be regarded as a buyer with the aim to maximize the difference between its gained transmission rate and the payment, while the IAB node aims at maximizing its revenue under a specific power allowance.
An iterative algorithm was proposed to obtain the price for the resource and the power allocation policy which reaches the state of proved unique Stackelberg equilibrium.

In a user-provided network (UPN), users can serve as providers, directly offering connectivity to other users ~ \cite{sofia2008user}. 
This architecture can provide users with good channel condition when the link to the BS suffers from poor quality, in a device-to-device (D2D) mode. 
However, in practice, users would have no incentive to be such a provider unless they receive satisfying rewards from the operator to compensate for their transmitting cost. 
By considering a D2D mode UPN under mmWave IAB HetNet, joint incentive and resource allocation design was studied in~\cite{Liu2020}. The authors formulated a Nash bargaining problem under the user utility, the sensitivity of battery energy, the incentive compensation, and the limitation of network resources constraints.
A centralized algorithm and a distributed algorithm were proposed to derive optimal Nash bargaining solution and decoupled sub-problem, respectively. 
The performance of the proposal is evaluated in terms of pay off, download data, and the operator's revenue.
It was shown that the joint optimization scheme can outperform the optimized resource or incentive only at least $16\%$ in every performance metric.

%%%%%%%%%%%%%%%%%%
\subsection{Network Coding-Aware Scheduling}
%With the usage of  degrees of freedom (DoF) as the performance metric, the authors in~\cite{Bande2020} investigated the DL transmission in a two-layered IAB network. 
%IAB donor 
%is studied in  with macro basestations (MBs), small-cell basestations (SBs) that act as half-duplex analog relays, and mobile
%terminals (MTs). 

Network coding enables the router to broadcast a mixture of information packets.
Due to its potential to provide energy-efficient and high throughput performance, network coding has found wider acceptance in the academia and industry~\cite{fragouli2006benefits,iqbal2011survey,magli2013network}.
%%%
For a HD hexagonal wireless backhaul network where the MBS transmits a linear combination of the messages to a set of SBSs, the per user degrees of freedom was shown to be approaching $1/2$ in~\cite{Bande2020}.
%%%% Thomsen2015
In~\cite{Thomsen2015}, the authors considered the integration of XOR network coding with the IAB network. 
It is assumed that the message sent by an IAB donor can be split into a common part and a private part.
An IAB node can XOR its decoded signal and broadcast the result to the mobile user and the IAB donor for DL and UL transmissions, respectively.  
With the aim to minimize the energy consumption of the IAB donor, 
the authors proposed the optimal policies for message splitting ratio, private part transmitting power, and common part beamformer design.
Meanwhile, the energy consumption minimization problem for the IAB nodes was formulated and solved by the convex optimization method.

The drawback of~\cite{Thomsen2015} is that only one-hop wireless backhaul is supportable and two mobile users are considered.
The properties of multi-hop backhaul and multi-route topology for the IAB network bring new challenges.
In~\cite{Mao2020}, with the consideration of more complex topology in IAB network, the authors proposed a linear network coding solution to enhance the performance robustness and reduce the latency.
A rate-proportional mechanism and an adaptive coded-forwarding mechanism were proposed to balance the traffic load among routes and determine the ratio of coded messages, respectively.
Compared with the round-robin mechanism, the rate-proportional mechanism can improve 25\%  spectral efficiency performance.
In comparison with ARQ retransmission, for a given input data rate, the adaptive coded-forwarding mechanism doubled the success rate.

\vspace{0.5em}
\subsection{Section Summary}
 This section mutually introduces and compares  literature for addressing the problems of network deployment and scheduling in IAB networks.
 Table~\ref{tab:TaxonomySec5} provides a taxonomy for the literature
according to the network model and the technique, the  majority  of  works focuses on multi-hop backhaul IAB networks.
The common drawback of all proposed solutions is that the mobility of the UEs is disregarded. In the network with a higher density of BSs, the movement of UEs will incur a more frequent need of handover, which brings additional challenges in the design of deploying and scheduling policies.

\section{Cache-enabled IAB Networks}\label{sec:6}
\begin{figure}
\centering
\includegraphics[width=0.6\columnwidth]{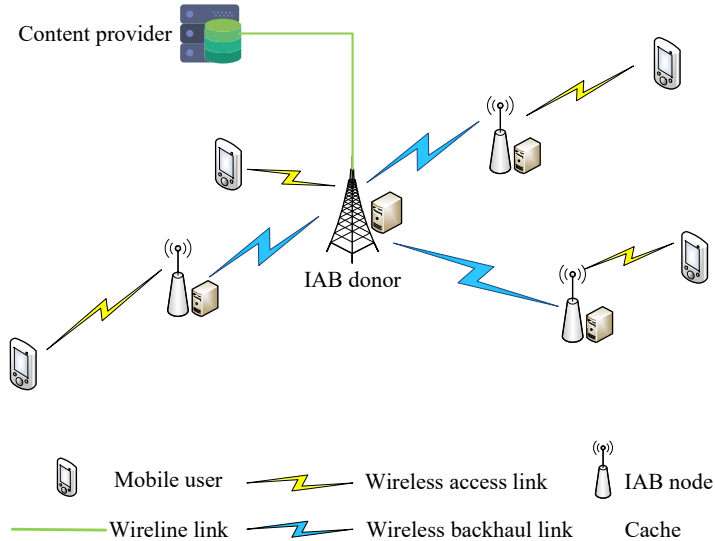}
\caption{Cache-enabled IAB network model.}
\label{fig:cache_network}
\end{figure}
Caching is able to improve the QoS experienced by users, reduce the overall network traffic, and prevent the network congestion~\cite{Yao2019}.
By caching popular contents from remote servers in network devices positioned closely to mobile users, the burden of wireless backhaul can be alleviated.
Fig.~\ref{fig:cache_network} provides a simple example of a cache-enabled IAB network, in which each BS with finite cache size can store some contents to serve mobile users.
 In this section, we survey the research focused on cache-enabled IAB network.

\vspace{0.5em}
\subsection{Successful Probability and Throughput Analysis}
The authors in \cite{Wang2018} considered a two-tier edge caching system, with limited storage resource SBSs and MIMO-enabled MBS.
The analytical framework for the successful content delivery probability (SCDP) was developed using tools from SG.
Further, a one-dimension search algorithm was proposed to obtain the minimum BSs density and maximum load at a small-cell under the SCDP requirement, respectively.
Simulation results revealed that the hit probability should be less than a certain level to achieve a given SCDP. 
The reason behind this requirement is that the larger hit probability could introduce more interference from SBS-tier and degrade the performance of cached content delivery.
Also, it was shown that the delay of non-cached content requests can be reduced close to the cached content request with the aid of MIMO-enabled backhaul.

With the consideration of using IBFD wireless backhaul, the authors investigated the average success probability (ASP) for a
cache-enabled mmWave IAB network in~\cite{Zhang2019}. 
The IAB nodes serve multiple users in IBFD mode with hybrid beamforming.
IAB nodes equipped with a storage memory to pre-cached some popular contents in order to alleviate network traffic loads. 
%The locations of IAB donors and IAB nodes are modeled by two independent Poisson point processes. 
The authors assumed a tuning parameter to control
the amount of RSI, and investigated two caching strategies: uniform (UC) and $M$ most popular files (MC), then derived the average rate, latency, and upper-bound for the average success probability (ASP) of file delivery.
In addition to investigating the latency and mean throughput, the authors derived a lower-bound for the ASP of content delivery. 
Simulation results showed that MC  significantly outperforms UC for ASP and the performance gap of rate and latency between IBFD and OBHD decreases as the storage space at IAB nodes increases.

In~\cite{Zhang2020}, Zhang \emph{et} \emph{al.} derived the average potential throughput (APT) of a two-tier out-band mmWave IAB cache-enabled network.
The IAB donors are connected to the core network via optical fiber links and contain all kinds of files. 
Similar to~\cite{Saha2018} and~\cite{Saha2019}, the total spectrum resource at IAB donor for the DL is partitioned into access and backhaul with a fixed ratio for the whole system. 
By using a SG tool, the authors developed an analytical framework to capture the SINR statistics and APT of users associated with each tier. 
Simulation results showed that the considered cache-enable setting can achieve 80\% APT improvement compared with the traditional setting. 

\vspace{1em}
\subsection{Caching Strategy Design}
%%%%%%
The cached content placement decision model is captured by fixed hit probability in~\cite{Wang2018,Zhang2020,Zhang2019}.
As suggested in~\cite{Wang2018} and~\cite{Zhang2019}, the hit probability plays an important role in the network performance.
%%%%%
In order to find the optimal content caching decision at IAB nodes and the mobile users association policy, the authors in~\cite{Haw2019} formulated a mixed-integer optimization problem with an objective function formulated as the weighted sum of transmission time and content hit number.
By using matching theory and Autoregressive Integrated Moving
Average (ARIMA) method, an iterative algorithm was proposed to solve the joint problem. 
%%%%
Compared with a conventional network, the proposed scheme brings up to 82\% performance improvement.
With the assumption that cache-enabled IAB nodes have three different backhaul options: 1) wired fiber, 2) mmWave wireless backhaul, and 3) sub-6 GHz wireless backhaul, authors proposed a game-theoretic learning algorithm to solve the content caching problem in~\cite{Hamidouche2017}. 
A minority game was formulated, where IAB nodes act as players with independent decisions for backhaul option choice and types of caching contents.  
In comparison with the greedy algorithm, the proposed scheme can achieve up to 85\% performance improvement.

%%%%%
Both \cite{Haw2019} and \cite{Hamidouche2017} assumed that the content popularity is stable over time, this assumption could lead to performance degradation in reality when the content popularity is unpredictable. 
In this context, with the consideration of dynamic content popularity, the authors in~\cite{Blasco2014} proposed a learning algorithm based on multi-armed bandit (MAB) theory to maximize the bandwidth alleviation for backhaul links.
The proposed algorithm is able to learn the content placement decision without the need for file popularity information.
In comparison with the random cache method, the proposed algorithm can achieve up to 7 times performance improvement.
%%%%
By introducing a virality parameter to represent the change of users' content requests over time, the authors in~\cite{Ahangary2020} studied the cached content placement problem in a mmWave IAB network.
Moreover, a hierarchical caching architecture was considered,  where the IAB donor and IAB nodes store different finite kinds of contents. 
The request for content stored in the central storage server will be forwarded by IAB donor via a wireless backhaul. 
In order to minimize backhaul load and average delay, the authors proposed a heuristic method to derive the optimal content placement decision.
Simulation results showed that the proposed method can achieve 
improvement for hit probability by 23\%, and reduction in backhaul load and average delay by 48\% and 27\%.

%%%%Gao2020
The coded cooperation of cache-enabled IAB nodes was considered in~\cite{Gao2020} and~\cite{Vu2018}.
The authors in~\cite{Vu2018} studied the energy-efficiency performance of a cache-enabled
IAB network, with the consideration of the cache capability. Simulation results showed that uncoded caching outperforms coded caching in the small user cache size regime, while coded caching outperforms uncoded caching in the small BSs cache size regime.
 In~\cite{Gao2020}, the authors applied the maximum distance separable (MDS) code at IAB nodes.
By leveraging the Q-learning model, the content caching decision is derived.
Compared with the DDPG method, the proposed method can reduce the time complexity significantly.	
The energy efficiency maximization problem in a FD cache-enabled IAB network was addressed in~\cite{Vu2020}.
 The authors investigated two collaboration schemes among the edge BSs : 1) distributed caching-separate access transmission and 2) cooperative caching-joint access transmission. The edge BSs are operated in FD mode. By optimal precoding design, numerical results showed that the cooperative caching scheme outperforms the distributed scheme.
\vspace{1em}
\subsection{Section Summary}
 The comparison of the studies related to the performance analysis and the caching strategy design for cache-enabled IAB networks is shown in Table~\ref{tab:TaxonomySec6}.
 The authors in~\cite{Wang2018} and~\cite{Zhang2019a} have proved that the hit probability has an important impact on the performance of cache-enabled IAB networks.
Most existing works aim at designing the optimal content placement policy, however, the  optimization for storage size is disregarded. In general, the BS with a larger storage size implies that more contents can be cached, and the hit probability will increase. The allocation of caching storages is still an open issue in the cache-enabled IAB networks.

\section{Optical IAB Networks}\label{sec:7}
Optical wireless communication (OWC) using wavelength band ranges from 350 nm to 1550 nm is a promising alternative solution to RF transmission~\cite{Ghassemlooy2015}.
This section reviews the studies for the incorporation of visible light communication (VLC) and free-space optical (FSO) communication into IAB networks, respectively. 
\vspace{1em}
\subsection{VLC}
In VLC, the transmission is based on the modulation of the intensity of the optical source, which uses the visible range of electromagnetic spectrum~\cite{Wu2020}.
Due to the energy-efficient benefit and long lifetime, Light-emitting diode (LED) is widely used in indoor illumination.
The authors in~\cite{Kazemi2017} developed an analytical framework to investigate SINR and average spectral efficiency in a hexagonal VLC-based IAB network.
Moreover, the effect of employing in-band and out-band wireless backhaul was studied. 
Compared with out-band backhaul, it was shown that the in-band mode can achieve better performance when the emission semiangle exceeds 25$^{\circ}$ or falls below 20$^{\circ}$.
The work in~\cite{Kazemi2017} is extended in~\cite{Kazemi2019} to a two-tier multi-hop wireless backhaul scenario.
For that setup, the authors derived the power control policy by solving the backhaul transmitting power consumption minimization problem.
It was shown that, in terms of average sum rate, a small value of the emission semiangle can achieve better performance, and in-band mode significantly outperforms out-band mode.

The joint backhaul spectrum resource and transmitting power allocation problem was addressed in~\cite{Kazemi2018}.
The authors first derived the analytical expressions for access and backhaul rate, and then used it to obtain the power and spectrum allocation coefficients under the backhaul capacity requirement.
From an optimization perspective, in~\cite{Kazemi2020}, based on projected subgradient method, with the aim to maximize the sum rate of mobile users and the IAB nodes,  user-centric and cell-centric backhaul spectrum scheduling policies were derived for the considered two-tier multi-hop VLC-based IAB network.
By integrating the RF transmission with VLC, the authors in~\cite{Murugaveni2020} considered a hybrid network with RF-based IAB donor and VLC-based IAB nodes.
In particular, the IAB node employed NOMA to improve spectral efficiency.
With the objective to minimize the interference, a heuristic algorithm was proposed to obtain the bandwidth reuse decision among the IAB nodes.
Simulation results showed that the proposed dynamic scheme can achieve performance improvement in both sum rate and spectrum efficiency.
\vspace{1.5em}

\subsection{FSO}
FSO enables high capacity transmission between access points (APs) that are separated by several kilometers, with a frequency that is above 300 GHz~\cite{Khalighi2014,Trichili:20}. 
Therefore, FSO communication is an attractive option for coping with the increased communication traffic demands especially in the outdoor environment~\cite{Alzenad2018}.
However, one of the inherent difficulties for FSO communication is the requirement for a reliable LoS. 
%%%%network reconfigurations
In~\cite{Gu2018a}, the authors first proposed a greedy algorithm to optimize the network sum rate and overall power cost, and to obtain the policies for connecting path and routing. 
Moreover, two network reconfiguration solutions to deal with the link failure and dynamic traffic demand changement were proposed, respectively.  
%%%%%Li2015a
In order to improve the reliability, the authors in~\cite{Li2015a} considered the usage of the mirror-aided links to ensure the connectivity of two distinct nodes when their LoS is blocked.
For the formulated graph optimization problem with the aim to minimize the sum of weighted cost and reliable link length, they proposed a sequential computation algorithm to obtain the connecting paths as well as the number and locations of mirrors, IAB donors, and FSO transceivers.
Based on a realistic topology model, the simulation results showed that the proposed method can achieve optimal or
near-optimal performance with much lower time complexity.

%%%%lj
The authors in~\cite{Atakora2018} considered an IAB network where the IAB donor is equipped with a FSO transceiver.
With the objective to minimize the overall delay under the requirement of all IAB nodes are covered, they solved the multicast problem for IAB nodes when they are static or mobile.
By reformulating the problem as a time-dependent prize collecting traveling salesman problem, the authors proposed several heuristic algorithms to find the optimal scheduling of the directional optical links.
%%%%% 
A hybrid RF/FSO UL transmission for IAB networks was studied in~\cite{Jamali2016}, where there are two types of links in the wireless backhaul.
In particular, the authors considered two transmission schemes at the IAB nodes: 1) delay-tolerant, where it can store the received information packets and forward them later; 2) delay-limited, where it forwards the received signal immediately.
By solving the throughput maximization problem via Lagrange dual decomposition method, the optimal UL block resource allocation policy was derived. 
Simulation results showed that sharing block resources is required to improve the throughput when the quality of FSO links falls below a certain level.
\vspace{2.5em}

\subsection{Section Summary}
  This section introduces and compares literature for the integration of IAB to the optical transport networks.
  The taxonomy for the literature according to the network model and the technique is shown in Table~\ref{tab:TaxonomySec7}. 
  In comparison to RF transport networks, the OWC system has a very high optical bandwidth resource. Hence, by exploiting the much larger bandwidth, it is less expensive to perform self-backhauling in the optical IAB networks than the RF counterparts. 
  As a possible non-3GPP research extension, in the optical IAB networks, the OWC links can be deployed between the BSs and the mobile switching center to enable a much higher backhaul throughput. As for the access links in the optical IAB networks, OWC links are only provided for the end users which can be equipped with optical transmitter and receiver, such as video-surveillance devices. The hybrid architecture consisting OWC and RF in the IAB networks can overcome the limitations of individual networks. Therefore, hybrid OWC/RF IAB networks are emerged as a promising solution for high-data-rate wireless communication systems.

\section{Non-Terrestrial IAB Networks}\label{sec:8}
Providing global connectivity is not a new ambition~\cite{Yaacoub_2020}.
To this end, satellite and unmanned aerial vehicle (UAV)
are attractive options to offer communication service over the areas that are too expensive to reach or too difficult to deploy conventional terrestrial networks (e.g., rural and remote regions, marine area)~\cite{Talgat2020,Kishk2020}.
This section focuses on the research related to the non-terrestrial IAB networks.
We classify these papers into two categories: (i) UAV-assisted IAB networks, and (ii) satellite-terrestrial IAB networks.
\vspace{0.5em}
\subsection{UAV-assisted IAB Networks}
%%%%Gapeyenko2018
Due to the ability to their mobility and relocation flexibility, UAV-BSs could be deployed at any 3D position of interest. 
However, due to the non-negligible multipath propagation and link blockage in a realistic environment, one important issue in a UAV-assisted network is the optimization of UAV-BSs positions.
In~\cite{perez2019ray}, based on ray-tracing simulations, the authors first investigated the coverage gains of UAV-assisted mmWave IAB network, where the UAV-BSs act as IAB nodes.
With the consideration of amplify-and-forward (AF) OBFD and decode-and-forward (DF) IBFD modes, the optimal positions for UAVs were derived via the ray tracing-based coverage maps.
The simulation results showed that the AF mode outperforms DF mode with 31\% DL coverage gains improvement.

Zhang \emph{et} \emph{al}. addressed the joint optimization problem for the number and three-dimensional (3D) positions of UAV-BSs in an IBFD UAV-assisted IAB network~\cite{Zhang2019b}. 
A heuristic algorithm was proposed to solve the optimization problem which aims to minimize the number of UAVs while maximizing the overall transmission rate.
Numerical results proved that the proposed algorithm can not only increase the overall throughput but also can decrease data rate block ratio.
In~\cite{Kalantari2017}, the authors solved the UAV-BSs placement problem in order to maximize the sum rate and the number of served mobile users. 
Moreover, the effect of the movement of mobile users was investigated in the simulations. 
Simulation results showed that the proposed method can achieve robust performance under the impact of users' mobility.
%%%%% Cicek2020
By defining a piecewise-linear function of gained transmission rate that captures the profit of mobile users, 
the authors in~\cite{Cicek2020} addressed the joint UAV-BS position and bandwidth allocation optimization problem.
In order to maximize the total network profit, they proposed a heuristic algorithm to solve the formulated mixed-integer non-linear programming problem.
Numerical results showed that the proposed solution is able to significantly outperform the well-known BARON solver \cite{Tawarmalani2004} and network-enabled optimization system (NEOS) platform\cite{Czyzyk1998}.

%%%%Dynamic
In a realistic environment, the dynamic change for both the multipath propagation and link blockage are non-negligible.
The authors in~\cite{Gapeyenko2018} developed analytical frameworks to characterize the performance of a UAV-assisted mmWave IAB network with terrestrial BSs (TBSs) and UAV-BSs.
With the consideration of the dynamic mobility of human blockers and UAV-BSs, the time-averaged and time-dependent performance metrics were derived.
It was shown that both outage probability (OP) and spectrum efficiency were improved as the intensity of UAV-BSs traversals increases. 
In addition, it was also shown that the lower flight speed can achieve a better wireless backhaul performance.
%%%%
The authors in~\cite{Zhang2020d} considered a maritime
communication system with cache-enabled UAV-BSs.
The optimal horizontal positions of UAV-BSs were obtained by an iterative one-dimensional linear search algorithm to maximize the sum rate. 
%%%% Khuwaja2020
In~\cite{Khuwaja2020}, the authors investigated the incorporation of cache-enabled UAV-BSs into terrestrial cache-enabled IAB network.
A tractable model for characterizing SCDP and energy efficiency was developed. 
Simulation results showed that the considered scheme can achieve 26.6\% SCDP performance improvement on average.

%%% Path selction
The path selection problem for a UAV-assisted multi-hop IAB network was addressed in~\cite{Almohamad2018}.
The authors formulated and solved a binary linear optimization program aiming at maximizing the total transmission rate, and obtain the optimal path scheduling policy.
By integrating UAVs as drone BSs into the IAB network, 
Fouda \emph{et} \emph{al}.~\cite{fouda2018uav} presented a system model for forward link transmissions in an  in-band IAB  HetNet. 
Given that the backhaul link to UAV is provided by the IAB donor,  an optimization problem is formulated that aims to achieve the maximum sum rate of the users, under the constrains that the mutual interference between access and backhaul links is below a given threshold. 
By using an alternative optimization method, the authors derived the optimal 3D hovering positions of the UAVs, mobile users association policy, precoder design at the backhaul link, and transmitting power allocation policy. Further, a better performance algorithm based on particle swarm optimization (PSO) is proposed in~\cite{Fouda2019}.

\vspace{0.5em}
\subsection{Satellite-Terrestrial IAB Networks}

%\begin{figure}
%\centering
%\includegraphics[width=0.8\columnwidth]{Sattmodel.png}
%\caption{UAV-assisted IAB system connectivity example.}
%\label{Satemodel}
%\end{figure}
One advantage of satellite-terrestrial communication is that it enables the network operators to cover wider areas at a lower cost. 
There are several on-going satellite-terrestrial communication projects, such as SpaceX and OneWeb, aiming to provide global-coverage and high data rates via  low earth orbit satellite (LEO) networks. 
In this context, the design of satellite-terrestrial backhaul link plays a key role in the service quality of users~\cite{Artiga2018}.

%Recently, a project named Shared Access terrestrial-satellite backhaul Network enabled by Smart Antennas (SANSA), provided a reliable solution for the backhaul to cope with the rapidly increasing traffic volumes, which is funded by the EU under the H2020 program.
%In SANSA, the researchers regard the satellite as an IAB node with fixed wireless backhaul link to the terrestrial system, while the terrestrial IAB nodes topology are reconfigurable and controlled bt the network manager to cope with the dynamically changing requirements for backhaul.

The authors in~\cite{Lagunas2017a} addressed the spectrum resource allocation problem for the DL transmission in a satellite-terrestrial IAB network.
The considered system assumes the terrestrial links and satellite links can reuse the same frequency.
By decomposing the sum rate maximization problem into two sub-problems, the optimal carrier allocation strategy was derived by the Hungarian method sequentially. 
%%% matching theory 
In \cite{Di2018}, a satellite-terrestrial IAB network architecture for data offloading was studied.
The satellite-terrestrial backhaul links share the Ka-band while the terrestrial links share the C-band. 
Terrestrial mobile users have access to the core network through the macro-cell, the small-cell, or the LEO-based small-cell.
With the aim to maximize the sum rate, the authors proposed a modified swap matching algorithm to obtain the optimal policies for users association, subchannel allocation, and power control. 
Simulation results revealed that the total capacity is not a monotonic function of the projected area of the satellite, which means that there exists an optimal satellite deployment solution to maximize the total backhaul capacity.

With the application of the game theory model into a satellite-terrestrial IAB network, the authors in~\cite{Deng2020} proposed a data offloading pricing mechanism based on the Stackelberg game model.
To deal with the increasing backhaul demand at conventional APs, they assumed some user's demand from conventional APs can be migrated to LEO-based APs.
In this context, the LEO-based APs act as the leaders while the conventional APs act as the followers.
An iterative algorithm was proposed to reach the Stackelberg equilibrium.
In each iteration, the follower level optimization problem aims at optimizing user association is solved by fractional programming, then the leader level problem with the aim to optimize the service price and Ka-band spectrum allocation is solved by alternative optimization.
The simulation results showed that there is an optimal LEO satellite density to balance the trade-off between APs utility and cost.  

In addition to the application of the game theory model, the integration of UAV and satellite-terrestrial IAB networks was investigated in~\cite{Hu2020}.
%% HAP 
%
The IAB nodes consist of the satellite, terrestrial SBSs, and UAVs, while the IAB donor is the MBS. 
This joint backhaul and access resource management problem was formulated as a competitive market.
Hu~\emph{et}~\emph{al.}~assumed that some UAVs, satellite, some SBSs, and MBSs act as ``sellers'', where others UAVs and SBSs act as ``customers''. 
The communication services were regarded as ``goods'' and their prices are determined by the QoS, while the cost is determined by the power consumption. 
The network seeks for achieving the Walrasian equilibrium, at which there are no good exits, and each role's profit is maximized. 
The simulation results showed that the proposed algorithm can reach Walrasian equilibrium within 200 iterations. 
Compared to random allocation, the proposed algorithm achieved 3 to 4 times gain in thedata rate.
 
\vspace{0.5em}

\subsection{Section Summary}
 The comparison of the studies focusing on the non-terrestrial IAB networks is shown in Table~\ref{tab:TaxonomySec8}. 
 The related studies aim at achieving economic wide-area connectivity, which can be classified into UAV-assisted and satellite-terrestrial IAB Networks. 
%%%%
 Non-terrestrial networks and IAB networks have been defined by 3GPP separately~\cite{3gppntn,3gpp15}. However, the integration of these two technologies is not standardized yet.
 The integration of non-terrestrial networks and IAB emerges as a promising solution for providing service in the area where optic fiber cables are not available. Notably, the performance of non-terrestrial IAB networks is affected by some individual characteristics of non-terrestrial networks, such as long propagation distance and movement of aerial BSs. As a result, the network deployment design will be an important issue in the non-terrestrial IAB networks.
%%%%%  Matching Game all rewr
%The cooperation of LEO satellite and high altitude platform (HAP) for providing communication service to remote regions was studied in~\cite{Jia2020}. 
%In order to maximize the weighted sum of successful connected users, Jia~\emph{et}~\emph{al.}~first formulated a 
%three-sided matching problem.
%For the tractable purpose,  a two-tier matching algorithm was proposed to address the dynamic connection between satellite and HAP due to the periodic movement of satellite.

%By incorporating laser-driven adaptive wireless-power-transfer at the UAV,  joint connectivity maximization and resource allocation problem in a RF powered backscatter communication  access link and free space optical backhaul link system was studied in \cite{Hassan2020}.
%An algorithm of polynomial computational complexity based on  three-stage optimization  is devised to solve the optimization problem.
%Using extensive simulations, efficiency of the proposed
%algorithm is demonstrated for improving the supportable arrival
%rate per IoT device and the number of the connected IoT devices
%in UL of showed-IoT network.

\section{Challenges and Opportunities}\label{sec:9}
The existing literature presented various promising solutions to improve the performance of the IAB networks.
In order to exploit the full potential of IAB networks, the flexible integration of IAB networks  with other advanced techniques may achieve even more success in the next generation communication system. 
The OWC links can be applied into the backhaul links in the cache-enabled and non-terrestrial IAB networks to improve the backhaul throughput, and the non-terrestrial IAB networks can adopt cache-enabled BSs to reduce the probability of requesting content from the remote server. 
For example, in a non-terrestrial IAB network with multiple drone BSs, we can consider adopting FSO links for the drone BS to drone BS links in order to achieve a higher data rate and reduce the inter-interference.
In particular, we discuss the challenges and possible opportunities
on the deployment, scheduling, mobility management, and intelligent
\vspace{1em}

\subsection{IAB Deployment}
In reality, the height of the blockages and the terrain information cannot be negligible for the UHF channel. 
Therefore, the 3D positions should be considered in the IAB deployment.
Moreover, the interaction among different layers of the protocol stack remains a design challenge for IAB.
Thus, designing a more realistic and efficient deployment solution for improving the end-to-end performance is still worthy research.
\vspace{1em}

\subsection{Scheduling}
DAG and ST limit the flexibility and efficiency due to the fixed parent-to-child relation between two adjacent IAB nodes. 
Studies in both \cite{Zhai2020} and \cite{islam2018investigation} proved that mesh-architecture based IAB achieves a considerable improvement in comparison to DAG-architecture based IAB.
It is expected that mesh-architecture-based IAB will be introduced in the Rel-17.
The solution to address congestion control and routing will be important in mesh-architecture-based IAB.
\vspace{1em}

\subsection{Mobility Management}
Due to the usage of higher frequencies, the wireless backhaul link in IAB is vulnerable to mobility blockage (e.g., vehicles and human movements). 
Therefore, from a resilience perspective, it is desired to ensure that an IAB network can provide reliable services to end-users when some backhaul links are degraded or lost.
 
\vspace{1em}

\subsection{Intelligent IAB}
In IAB, the relation between access and backhaul will be closer than ever.
Therefore, joint operation is require since total separation of their resources might not be possible.
A crucial issue in operator maintenance is the designing of a mechanism for realizing an intelligent access-demand-aware backhaul system in which a central or distributed controller optimizes the backhaul capacity according to dynamic access network demand.
\vspace{0.5em}

\section{Conclusions}\label{sec:10}
IAB provides an economical and flexible network deployment solution in the era of 5G and beyond network densification.
In this study, a brief introduction to IAB was followed by
 a discussion and comparison of recent research works on IAB networks.
In addition, state-of-the-art studies on integrated IAB with cache-enabled, optical transport, and non-terrestrial communication networks were reviewed. 
Finally, the possible challenges and opportunities in this promising research area were described.

\section*{Author Contributions}
The work was developed as a collaboration among all authors.
All authors conceived the work and suggested the
outline of the paper. The manuscript was mainly drafted by YZ
and was revised and corrected by all co-authors. 
All authors have
read and approved the final manuscript.

\section*{Acknowledgement}
This work was supported by the Office of Sponsored Research
(OSR) at the King Abdullah University of Science and Technology
(KAUST).

\section*{Tables}
%%%% Table sec3
\begin{small}
\begin{center}
\begin{longtable}{|c|c|c|}
\caption{Taxonomy of Literature in Section~\ref{sec:3}.}\label{tab:TaxonomySec3}
\\\hline
   \textbf{Objective}  & \textbf{Network} & \textbf{Reference} \\\hline
   \endfirsthead
\multicolumn{3}{c}%
{\tablename\ \thetable\ -- \textit{Continued from previous page}} \\
\hline
\textbf{Objective} & \textbf{Network} & \textbf{ Reference} \\
\hline
\endhead
\hline \multicolumn{3}{r}{\textit{Continued on next page}} \\
\endfoot
\hline
\endlastfoot
 \hline  
 %%%%% SG analysis
     { \makecell*[c]{Coverage Probability}} 
 & \makecell*[c]{Multi macro-cells, \\TDD, mmWave}  &  \cite{Kulkarni2017}  \\\cline{2-3}       
    &   \makecell*[c]{hybrid UHF  \& mmWave, \\OBHD }  & \cite{Singh2015}  \\\cline{2-3}         
   &  \makecell*[c]{Multi macro-cells, \\IBFD} &   \cite{Sharma2017} \\ \cline{2-3}  
   &  \makecell*[c]{Single macro-cell, \\OBHD, mmWave} & \cite{Saha2018} \\ \cline{2-3}  
   &  \makecell*[c]{Multi macro-cells, \\OBHD, mmWave}&  \cite{Saha2019} \\ \cline{2-3} 
   & \makecell*[c]{massive MIMO, OBFD} &  \cite{Tabassum2015} \\   \cline{1-1}\cline{2-3}
 \makecell*[c]{Ergodic Capacity}  & \makecell*[c]{Multi macro-cells,\\ IBFD} &   \cite{Zhang2018} \\   \hline 
\end{longtable}
\end{center}
\end{small}

\begin{small}
\begin{center}
\begin{longtable}{|c|c|c|c|c|}
\caption{Taxonomy of Literature in Section~\ref{sec:4}.}\label{tab:TaxonomySec4}
\\\hline
   \textbf{Objective} & \textbf{Methodology} & \textbf{Network} & \textbf{\makecell*[c]{Optimization\\ parameters} } & \textbf{Reference} \\\hline
   \endfirsthead
\multicolumn{5}{c}%
{\tablename\ \thetable\ -- \textit{Continued from previous page}} \\
\hline
\textbf{Objective}  & \textbf{Methodology} & \textbf{Network} & \textbf{\makecell*[c]{Optimization\\ parameters} } & \textbf{ Reference} \\
\hline
\endhead
\hline \multicolumn{5}{r}{\textit{Continued on next page}} \\
\endfoot
\hline
\endlastfoot
%%%% RA
   {\makecell*[c]{Interference \\ Mitigation}}                         
   & Random matrix & \makecell*[c]{massive MIMO,\\ IBFD} &
   \multirow{3}{*}{\makecell*[c]{ Beamforming\\ matrix}}  & \cite{Li2015}\\ \cline{2-3}\cline{5-5}
   %%%%%    
     & Heuristic method &\makecell*[c]{ MIMO,\\ IBFD}& &\cite{Ullah2016}\\ \cline{2-3}\cline{5-5}
    %%%%%
     & Convex optimization  & MIMO,TDD & &\cite{jayasinghe2020traffic}\\ \cline{2-5}
   %%%%%     
     &\makecell*[c]{Random matrix,\\Heuristic approach} & \makecell*[c]{mmWave,\\mesh\\ backhauling} &{\makecell*[c]{ Channel alignment,\\  power control,\\ node placement}} & \cite{Nakamura2019}\\
      \cline{2-5}   
      
%%%%% 
%   \multirow{4}{*}
   {\makecell*[c]{Energy\\ Consumption\\ Minimization}}       
   & Heuristic method & DRX & Time allocation & \cite{Prasad2017}\\ \cline{2-5}
   %%%   
     & Heuristic method &\makecell*[c]{ mmWave,\\ multi-hop \\backhaul} &
     {\makecell*[c]{ Time allocation, \\  power control }} & \cite{Meng2018}\\ \cline{2-5}   
     %%% 
     &\makecell*[c]{Convex optimization} & \makecell*[c]{massive MIMO\\ IBFD} & \multirow{2}{*}{\makecell*[c]{   Power control }}& \cite{Korpi2018}\\ \cline{2-3} \cline{5-5}
     %%%%%  
    & Alternative optimization  & NOMA, IBFD &  & \cite{Lei2018}\\ \cline{1-5} 
 %%%%% 
   {\makecell*[c]{Throughput\\ Maximization}}  
   & DRL & OFDMA &Bandwidth allocation & \cite{lei2020}\\ \cline{2-5}
   %%% 
   & Heuristic method  & TDD  & {\makecell*[c]{ Time allocation, \\  bandwidth allocation }}&  \cite{Liu2018a}  \\ \cline{2-5}
   %%%% 
     &  Water-filling algorithm & IBFD & \multirow{2}{*}{Power allocation} & \cite{Lagunas2017}   \\ \cline{2-3}\cline{5-5}
     %%%%% 
       & Random matrix & \makecell*[c]{MIMO,\\ IBFD } & &  \cite{Korpi2016}  \\ \cline{2-5}
       %%%%%
   & SCP & \makecell*[c]{mmWave,\\ IBFD} & {\makecell*[c]{ Power allocation, \\  bandwidth allocation }}&  \cite{Zhang2020b}  \\ \cline{2-5}
   %%% 
   &\makecell*[c]{Convex optimization,\\ bisection search} &  \makecell*[c]{hybrid IBFD \\ and OBHD} & Bandwidth allocation &\cite{Siddique2017}   \\ \cline{2-5}
   %%%% 
   & Game theory & mmWave & {\makecell*[c]{ Time allocation, \\  power control }}&  \cite{Liu2018}  \\ \cline{2-5}
   %%%
   & Game theory &  OBFD & {\makecell*[c]{ Sub-carrier\\ allocation }} &  \cite{Lashgari2017}  \\ \cline{2-5}
   %%% 
   & \makecell*[c]{Game theory,\\DRL} & IBFD &Power control &  \cite{Blasco2013}  \\ \cline{2-5}
 & \makecell*[c]{Random matrix,\\Alternative optimization}  &\makecell*[c]{ mmWave,\\ hybrid\\ beamforming} &{\makecell*[c]{ Power control,\\
 time allocation,\\
 user association,\\
 beamforming matrix }} &  \cite{Kwon2019}  \\ \cline{2-5}
 %%%%%
   & Random Matrix & \makecell*[c]{Hybrid\\  mmWave,\\Sub-6 GHz } &Beamforming matrix & \cite{Ni2019}   \\ \cline{2-5}
   %%%
   & Markov approximation & OBFD  &  {\makecell*[c]{ Sub-carrier\\ allocation }} & \cite{Pu2019}   
   \\ \cline{2-5}
       %%%%
   & CCP & NOMA & Power control &   \cite{Yang2020}  \\ \cline{2-5}
      %%% 
   & Riemannian optimization & \makecell*[c]{IBFD, \\ full CSI} &  \multirow{2}{*}{\makecell*[c]{Beamforming matrix,\\SBSs clustering }}& \cite{Chen2018}  \\ \cline{2-3}\cline{5-5}
   %%%
   & SCP  &  \makecell*[c]{IBFD, \\ partial CSI} & & \cite{Chen2019}  \\ \cline{1-5}
  %%%%%  
   \multirow{4}{*}{\makecell*[c]{Utility \\Optimization }}
  &  Lyapunov optimization & mmWave &\makecell*[c]{User association,\\ beamforming matrix,\\power control} &  \cite{vu2017joint} \\ \cline{2-5}
   %%%%
    & \makecell*[c]{ RL,\\Lyapunov optimization }& \makecell*[c]{mmWave,\\ multi-hop\\ backhaul} &  \makecell*[c]{Rate allocation,\\user association}  &  \cite{Vu2019} \\ \cline{2-5}
    %%%%% 
  & Alternative optimization & \makecell*[c]{Network } & \multirow{2}{*}{\makecell*[c]{Bandwidth allocation,\\user association}} &  
 \cite{Chen2016a} \\ \cline{2-2}\cline{5-5}
  & Lyapunov optimization & \makecell*[c]{ virtualization} &  & \cite{Tang2018} \\ \cline{1-5}
  %%%% 
  \makecell*[c]{EE \\ Maximization} & CCP & \makecell*[c]{massive MIMO,\\  EH, IBFD} & {\makecell*[c]{Beamforming mateix, \\power control,\\user association}}& \cite{Chen2016c} \\ \cline{1-5}
  %%%%% 
     \multirow{2}{*}{\makecell*[c]{SE \\ Maximization}} & SQP &\makecell*[c]{OBFD,\\Self-organizing\\ network} & \makecell*[c]{Antenna tilt\\ angle} & \cite{Imran2014} \\\cline{2-5}
      & SCP & \makecell*[c]{IBFD}& Power control & \cite{Chen2016b} \\
 \hline   
\end{longtable}
\end{center}
\end{small}

\begin{small}
\begin{center}
\begin{longtable}{|c|c|c|c|}
\caption{Taxonomy of Literature in Section~\ref{sec:5}.}\label{tab:TaxonomySec5}
\\\hline
   \textbf{Objective} & \textbf{Methodology} & \textbf{Network} & \textbf{Reference} \\\hline
   \endfirsthead
\multicolumn{4}{c}%
{\tablename\ \thetable\ -- \textit{Continued from previous page}} \\
\hline
\textbf{Objective} & \textbf{Methodology} & \textbf{Network} & \textbf{ Reference} \\
\hline
\endhead
\hline \multicolumn{4}{r}{\textit{Continued on next page}} \\
\endfoot
\hline
\endlastfoot
 \hline  
 \multirow{6}{*}{\makecell*[c]{Deployment\\ Design}}                          & Heuristic method & \makecell*[c]{massive MIMO,\\ad-hoc} & \cite{Bonfante2018}\\ \cline{2-4}
        & Convex optimization & \makecell*[c]{OFDMA,\\ MIMO} & \cite{Lai2020}\\ \cline{2-4} 
   & Heuristic method  & \multirow{2}{*}{\makecell*[c]{mesh\\ multi-hop \\backhaul,\\ mmWave}} & \cite{Wainio2016}\\ \cline{2-2}\cline{4-4} 
   & \makecell*[c]{K-means clustering,\\Genetic algorithm\\ algorithm} &     & \cite{Raithatha2020}\\ \cline{2-4}
           & \makecell*[c]{branch-and-bound \\algorithm} &  \makecell*[c]{ multi-hop \\backhaul, \\mmWave} & \cite{islam2017integrated}\\ \cline{2-4}
             & Genetic algorithm & IBFD & \cite{Rezaabad2018}\\              \hline 
             %%%%% 
  {\makecell*[c]{Routing}}          
  &  \makecell*[c]{Virtual-network\\ method} & \makecell*[c]{mesh\\ multi-hop \\backhaul, \\mmWave} & \cite{Hui2013}  \\ \cline{2-4} 
  & Heuristic method & \makecell*[c]{mesh\\ multi-hop \\ in-band\\ backhaul,\\ mmWave} & \cite{Favraud2017}  \\ \cline{2-4} 
  &Convex optimization & \makecell*[c]{multi-hop \\backhaul,\\ mmWave}& \cite{islam2018investigation}  \\ \cline{2-4} 
  & Heuristic method &\makecell*[c]{dual-hop \\ backhaul,\\TDD, \\mmWave} & \cite{Lukowa2018}  \\ \cline{2-4} 
  & \makecell*[c]{Distributed greedy \\algorithm} & \makecell*[c]{multi-hop\\ backhaul,\\ mmWave}& \cite{polese2018distributed}  \\ \cline{2-4} 
  & Backpressure algorithm & {\makecell*[c]{multi-hop\\ backhaul,\\ mmWave\\ MU-MIMO}}& \cite{GomezCuba2020a}  \\ \cline{2-2} \cline{4-4}
  &\makecell*[c]{Simulated annealing\\algorithm} & & \cite{GomezCuba2020}  \\ \cline{2-4} 
  & DRL & \multirow{2}{*}{\makecell*[c]{multi-hop,\\ mmWave}} & \cite{Gupta2019}  \\ \cline{2-2} \cline{4-4}
  &\makecell*[c]{RL, CG \\method}  & & \cite{Zhang2020c}  \\ \cline{2-4} 
  &  \makecell*[c]{Semi-distributed\\ learning algorithm} & \makecell*[c]{Dynamic\\ mmWave \\network} & \cite{Ortiz2019}  \\ \cline{2-4} 
    &Heuristic algorithm  & mmWave & \cite{Chaudhry2020}  \\ \cline{1-4} 
    %%%%             
   \makecell*[c]{\makecell*[c]{Fairness}}& \makecell*[c]{Matching theory} & TDMA mmWave& \cite{Yuan2018}\\\cline{2-4}  
   &Heuristic method  &\makecell*[c]{hybrid \\IBFD\\\&OBHD} & \cite{Siddique_2015}\\\cline{2-4}          
   & WPF algorithm & \makecell*[c]{DAG\\ multi-hop \\backhaul}& \cite{zhang2020co}\\\cline{2-4}          
   & Backpressure algorithm & IBFD & \cite{Goyal2017}\\\cline{1-4}   
   %%%%
      {\makecell*[c]{Incentive\\ Mechanism }}& Game theory & \makecell*[c]{Dual-hop\\ backhaul} & \cite{Rahmati2015}\\\cline{2-4} 
   & Game theory & \makecell*[c]{UPN, \\multi-hop\\backhaul}& \cite{Liu2020}\\\cline{1-4} 
   {\makecell*[c]{Network \\Coding-
   \\Aware \\Scheduling }}& \makecell*[c]{Convex optimization} & \makecell*[c]{XOR network\\ coding} & \cite{Thomsen2015}\\\cline{2-4}       
   & Coding design & \makecell*[c]{Linear network\\ coding,\\ multi-hop\\ backhaul} & \cite{Mao2020}\\ \hline 
\end{longtable}
\end{center}
\end{small}

\begin{small}
\begin{center}
\begin{longtable}{|c|c|c|c|}
\caption{Taxonomy of Literature in Section~\ref{sec:6}.}\label{tab:TaxonomySec6}
\\\hline
   \textbf{Objective} & \textbf{Methodology} & \textbf{Network} & \textbf{Reference} \\\hline
   \endfirsthead
\multicolumn{4}{c}%
{\tablename\ \thetable\ -- \textit{Continued from previous page}} \\
\hline
\textbf{Objective} & \textbf{Methodology} & \textbf{Network} & \textbf{ Reference} \\
\hline
\endhead
\hline \multicolumn{4}{r}{\textit{Continued on next page}} \\
\endfoot
\hline
\endlastfoot
%  
%  \makecell*[l]{SE Maximization} & & & \cite{Imran2014,} \\ 
   \hline                       
   \makecell*[c]{SCDP\\Analysis}  &\multirow{3}{*}{\makecell*[c]{Stochastic geometry}} & {\makecell*[c]{massive MIMO,\\OBFD,\\Caching }} & \cite{Wang2018}\\ \cline{1-1}\cline{3-4}
  \makecell*[c]{ASP\\Analysis}    & &\makecell*[c]{ IBFD,\\ mmWave,\\Caching} &         \cite{Zhang2019} \\ \cline{1-1}\cline{3-4}
    \makecell*[c]{Throughput \\Analysis} & &\makecell*[c]{ OBHD,\\ mmWave,\\Caching} & \cite{Zhang2020}\\\cline{1-4}
    %%%%
 \multirow{4}{*}{\makecell*[c]{Caching\\Strategy\\ Design}}& \makecell*[c]{Matching theory,\\ ARIMA} & \makecell*[c]{Single-hop\\ backhaul}& \cite{Haw2019} \\\cline{2-4}
 & \makecell*[c]{MAB-based\\learning} & \makecell*[c]{Unknown content \\popularity} & \cite{Blasco2014} \\\cline{2-4}
 & \makecell*[c]{Game theory,\\ RL} &\makecell*[c]{ Hybrid\\ mmWave\\\&Sub-6 Ghz}& \cite{Hamidouche2017} \\\cline{2-4}
 & DRL & \makecell*[c]{MDS-coded\\ caching}& \cite{Gao2020} \\\cline{2-4}
 &Heuristic algorithm & mmWave & \cite{Ahangary2020} \\
  \hline   
\end{longtable}
\end{center}
\end{small}

\begin{small}
\begin{center}
\begin{longtable}{|c|c|c|c|}
\caption{Taxonomy of Literature in Section~\ref{sec:7}.}\label{tab:TaxonomySec7}
\\\hline
   \textbf{Objective} & \textbf{Methodology} & \textbf{Network} & \textbf{Reference} \\\hline
   \endfirsthead
\multicolumn{4}{c}%
{\tablename\ \thetable\ -- \textit{Continued from previous page}} \\
\hline
\textbf{Objective} & \textbf{Methodology} & \textbf{Network} & \textbf{ Reference} \\
\hline
\endhead
\hline \multicolumn{4}{r}{\textit{Continued on next page}} \\
\endfoot
\hline
\endlastfoot
 \makecell*[c]{Interference\\ Mitigation}  & Heuristic algorithm & \makecell*[c]{VLC\\ \& NOMA}& \cite{Murugaveni2020}\\\cline{1-4}
 \makecell*[c]{Power \\ Consumption \\ Minimization \\ \&\\
  SINR   Analysis }& \multirow{2}{*}{\makecell*[c]{Projected subgradient \\ \& Stochastic analysis}} & VLC & \cite{Kazemi2017,Kazemi2019} \\ \cline{1-1}\cline{4-4}
   \multirow{3}{*}{\makecell*[c]{Throughput\\ Maximization}}& & & \cite{Kazemi2018,Kazemi2020} \\ \cline{2-4}
   &\makecell*[c]{Lagrange dual\\ decomposition} & \multirow{4}{*}{FSO} & \multirow{2}{*}{\cite{Jamali2016}} \\ \cline{1-2}\cline{4-4}
    Routing & \multirow{2}{*}{Heuristic algorithm}  &  & \cite{Gu2018a} \\ \cline{1-1}\cline{4-4}
    \makecell*[c]{Delay \\ Minimization} & & & \cite{Atakora2018} \\ \cline{1-2}\cline{4-4}
  {\makecell*[c]{Deployment\\ Design}} & \makecell*[c]{ Sequential\\ computation} &  &\cite{Li2015a} \\
  \hline   
\end{longtable}
\end{center}
\end{small}

\begin{small}
\begin{center}
\begin{longtable}{|c|c|c|c|}
\caption{Taxonomy of Literature in Section~\ref{sec:8}.}\label{tab:TaxonomySec8}
\\\hline
   \textbf{Objective} & \textbf{Methodology} & \textbf{Network} & \textbf{Reference} \\\hline
   \endfirsthead
\multicolumn{4}{c}%
{\tablename\ \thetable\ -- \textit{Continued from previous page}} \\
\hline
\textbf{Objective} & \textbf{Methodology} & \textbf{Network} & \textbf{ Reference} \\
\hline
\endhead
\hline \multicolumn{4}{r}{\textit{Continued on next page}} \\
\endfoot
\hline
\endlastfoot
  \multirow{5}{*}{\makecell*[c]{Deployment\\ Design}}  & Ray tracing & \multirow{4}{*}{UAV-terrestrial} & \cite{perez2019ray} \\ \cline{2-2}\cline{4-4}
  & \multirow{3}{*}{Heuristic algorithm} & & \cite{Zhang2019b}\\\cline{4-4}
  & & & \cite{Kalantari2017} \\\cline{4-4}
    & & & \cite{Cicek2020} \\\cline{2-4}
   &Linear search & {UAV-marine}&\cite{Zhang2020d} \\\cline{1-4}
  \makecell*[c]{SCDP\\ Analysis} &  \multirow{2}{*}{Stochastic geometry} & \makecell*[c]{Cache-enabled\\ UAV} & \cite{Khuwaja2020} \\\cline{1-1}\cline{3-4}
  \makecell*[c]{OP \\ Analysis} & & \multirow{3}{*}{UAV-terrestrial} & \cite{Gapeyenko2018}
 \\\cline{1-2}\cline{4-4}
  \multirow{2}{*}{Routing}  & Linear programming &  & \cite{Almohamad2018}
 \\\cline{2-2}\cline{4-4}
   & Heuristic algorithm &  & \cite{Fouda2019,fouda2018uav} \\\cline{2-4}
 \makecell*[c]{Throughput \\Maximization}  & Hungarian method & \multirow{2}{*}{Satellite-terrestrial}  & \cite{Lagunas2017a}
 \\\cline{2-2}\cline{4-4}
   & Swap matching  &  & \cite{Di2018}\\\cline{1-4}
 \makecell*[c]{Incentive\\ Mechanism} & Game theory & \makecell*[c]{Satellite-aerial-\\terrestrial}& \makecell*[c]{\cite{Deng2020}\\ \cite{Hu2020}}\\  \hline   
\end{longtable}
\end{center}
\end{small}

\bibliographystyle{IEEEtran}
\bibliography{IEEEabrv,ref_survey}

\end{document}